\documentclass[11pt,amsfonts,amssymb,amscd,xy,eufrak]{article}

\usepackage{amssymb}
\usepackage{amsmath}
\usepackage[all]{xy}
\usepackage{eufrak}
\newcommand{\dst}{\displaystyle}

\newcommand{\CC}{{\mathbb C}}
\newcommand{\RR}{{\mathbb R}}
\newcommand{\ZZ}{{\mathbb Z}}
\newcommand{\PP}{{\mathbb P}}

\renewcommand{\SS}{{\mathbb S}}
\newcommand{\End}{{\rm End}}
\newcommand{\Ker}{{\rm Ker}}
\newcommand{\Tr}{{\rm Tr}}
\newcommand{\diag}{{\rm diag}}
\newcommand{\ra}{\rightarrow}
\newcommand{\hE}{{\hat E}}
\newcommand{\hA}{{\hat A}}
\newcommand{\hphi}{{\hat \phi}}
\newcommand{\hX}{{\hat X}}
\newcommand{\hx}{{\hat x}}
\newcommand{\hF}{{\hat F}}
\newcommand{\bz}{\bar{z}}
\newcommand{\bs}{\bar{s}}
\newcommand{\hL}{\hat{L}}
\newcommand{\hD}{\hat{D}}
\newcommand{\hS}{\hat{S}}
\newcommand{\ha}{\hat{a}}
\newcommand{\hP}{\hat{P}}
\newcommand{\hQ}{\hat{Q}}
\newcommand{\bpartial}{\bar{\partial}}
\renewcommand{\lg}{\langle}
\newcommand{\rg}{\rangle}
\newcommand{\ind}{{\rm Ind}}
\newcommand{\pchi}{{\partial_\chi}}

\newcommand{\hDd}{\hat{D}^{\dagger}}
\newcommand{\ot}{\otimes}
\newcommand{\bD}{{\bar{\cal D}}}
\newcommand{\tw}{{\tilde{w}}}
\newcommand{\hK}{{\hat{K}}}
\newcommand{\chE}{\check{\hE}}
\newcommand{\hSS}{{\hat \SS}}
\newcommand{\cP}{{\mathcal P}}
\newcommand{\cA}{{\mathcal A}}
\newcommand{\hpi}{{\hat \pi}}
\newcommand{\ups}{{\mathfrak v}}

\newcommand{\Dd}{D^{\dagger}}
\newcommand{\dbz}{d\bar{z}}

\newcommand{\hLambda}{\hat{\Lambda}}
\newcommand{\hdelta}{\hat{\delta}}
\newcommand{\C}{{\bf C}}
\newcommand{\tC}{{\tilde{\bf C}}}
\newcommand{\Sc}{{\bf S}}
\renewcommand{\S}{S}

\newcommand{\E}{{\cal E}}
\newcommand{\CO}{{\cal C}}

\newcommand{\g}{\EuFrak{g}}

\newcommand{\Ebl}{\EuFrak{E}}

\newcommand{\nn}{\nonumber}
\newcommand{\tE}{\widetilde{E}}

\newcommand{\Ah}{\hat{A}}
\newcommand{\phih}{\hat{\phi}}
\newcommand{\Psih}{\hat{\Psi}}
\newcommand{\Psid}{\Psi^{\dagger}}
\newcommand{\Dhd}{\hat{D}^{\dagger}}
\newcommand{\Psihd}{\hat{\Psi}^{\dagger}}
\newcommand{\phihd}{\hat{\phi}^{\dagger}}
\newcommand{\G}{(\Dd D)^{-1}}
\newcommand{\Xh}{\hat{X}}

\newcommand{\Dh}{\hat{D}}
\newcommand{\hpsi}{\hat{\psi}}

\title{\bf Nahm Transform For Periodic Monopoles\\ And\\ {\cal N}=2 Super Yang-Mills Theory}
\author{
Sergey Cherkis\thanks{email: cherkis@physics.ucla.edu}\\
\it TEP, UCLA Physics Department, Los Angeles, CA 90095-1547
 \rm
\and 
Anton Kapustin\thanks{email: kapustin@ias.edu}\\
\it Institute for Advanced Study, Olden Lane, Princeton, NJ 08540
}

\begin{document}

\begin{titlepage}

\maketitle

\begin{sloppypar} 
\begin{abstract} 
We study Bogomolny equations on $\RR^2\times \SS^1$. Although
they do not admit nontrivial finite-energy solutions, we show that there are interesting
infinite-energy solutions with Higgs field growing logarithmically
at infinity. We call these solutions periodic monopoles. Using Nahm
transform, we show that periodic monopoles are in one-to-one correspondence
with solutions of Hitchin equations on a cylinder with Higgs field growing 
exponentially at infinity. The moduli spaces of periodic monopoles belong to
a novel class of hyperk\"ahler manifolds and have applications to
quantum gauge theory and string theory. For example, we show that the moduli
space of $k$ periodic monopoles provides the exact solution of ${\cal N}=2$ super
Yang-Mills theory with gauge group $SU(k)$ compactified on a circle of arbitrary 
radius.

\end{abstract} 

\vspace{-8.5in}

\vspace{1in}\parbox{\linewidth}
{\small\hfill \shortstack{IASSNS-HEP-00/44\\ UCLA-00TEP-17 \\ CITUSC/00-25}} 
\vspace{8.5in}

\end{sloppypar}
\end{titlepage}

\section{Introduction and Summary}\label{sec:intro}

\subsection{The Bogomolny Equation}\label{one}

Let $X$ be a three-dimensional oriented Riemannian manifold, $E$ be a vector 
bundle over $X$ with structure group $G$, $A$ be a connection on $E$, and
$\phi$ be a section of $\End(E)$. 
Bogomolny equation is the reduction of the self-duality equation to three 
dimensions which reads
\begin{equation}\label{Bogomolny}
F_A=*\,d_A\phi.
\end{equation}
Here $*$ is the Hodge star operator. In what
follows we set $G=SU(2)$, with $E$ being associated with the fundamental
representation of $SU(2)$. In this case $\phi$ is Hermitian and 
traceless. We will assume that all functions and connections
are infinitely differentiable, unless specified otherwise.

It is well known that for $X=\RR^3$ with flat metric the Bogomolny equation 
admits finite energy solutions, so-called BPS monopoles. The energy and magnetic charge of a 
pair $(A,\phi)$ are defined as follows:
\begin{eqnarray}
E(A,\phi)& = &\frac{1}{4}\int_X\ \Tr\left(F_A\wedge *F_A+*\,d_A\phi\wedge d_A\phi\right),\nn\\
m(A,\phi)& = &\lim_{R\ra\infty} \int_{|x|=R} \frac{\Tr\left( F_A\phi\right)}{4\pi||\phi||}.\nn
\end{eqnarray}
Here $$||\phi||^2=\frac{1}{2}\Tr\ \phi^2.$$ 

For BPS monopoles $||\phi||$ tends to a constant value $v$ at infinity, while
$||F_A||$ decreases as $1/r^2$. It follows that
the energy of a BPS monopole is proportional to its magnetic charge:
\begin{equation}
E(A,\phi)=2\pi\ v\ m(A,\phi).\nn
\end{equation}

BPS monopoles are absolute minima of the energy function $E(A,\phi)$
in a subspace with fixed magnetic charge and fixed asymptotic value of $||\phi||$.

\subsection{Periodic Monopoles}

Solutions of the Bogomolny equation on $\RR^2\times \SS^1$ have not
been studied previously. One of the reasons is that any monopole on 
$\RR^2\times \SS^1$ with a nonzero magnetic charge must have infinite
energy. This happens because the magnetic field of
a magnetically charged object on $\RR^2\times \SS^1$ decays only
as $1/r$, where $r$ is the radial distance on $\RR^2$. Hence the magnetic
energy density decays as $1/r^2$, and its integral over $\RR^2\times \SS^1$ 
diverges.

Since the energy of a periodic monopole is infinite, it cannot be regarded
as a solitonic particle. Still, periodic monopoles do play a role in certain
physical problems. For example, we will see that the centered moduli space of 
an $SU(2)$ 
periodic monopole with magnetic charge $k$ is a hyperk\"ahler manifold of dimension 
$4(k-1)$. It turns out that this hyperk\"ahler manifold coincides with the 
quantum Coulomb branch of the ${\cal N}=2$ super Yang-Mills theory on $\RR^3\times \SS^1$ with
gauge group $SU(k)$ (see below). For $k=2$ this manifold is a very interesting
deformation of the Atiyah-Hitchin manifold (the reduced moduli space of
a $k=2$ monopole on $\RR^3$) and is an example of a new class of
asymptotically locally flat self-dual gravitational instantons. The properties
of the moduli spaces of periodic
monopoles will be discussed in more detail in a forthcoming paper~\cite{usthree}. 

The goal of this paper is two-fold. On one hand, we want to compute the dimension of 
the moduli spaces of periodic monopoles and to establish a correspondence between periodic monopoles and 
solutions of Hitchin equations on a cylinder. The latter correspondence is a particular 
instance of Nahm transform.
On the other hand, we want to explain the relation of our results to the
four-dimensional ${\cal N}=2$ gauge theories
and to the brane configurations of the type first considered by Chalmers and 
Hanany~\cite{ChH} and further explored by Hanany and Witten~\cite{HW},
Witten~\cite{Witten}, and many others. These brane configurations were used to find the exact 
Coulomb branch of the super Yang-Mills theory with eight supercharges on
$\RR^3$ and $\RR^4$. In particular, Witten showed how to obtain the exact
solution of the ${\cal N}=2$ super Yang-Mills on $\RR^4$ from the physics of the M-theory 
fivebrane. As explained below, the Nahm transform approach not only 
reproduces the classical physics of the fivebrane, but goes considerably further by 
resumming the effects of membrane instantons. 

In a companion paper~\cite{ustwo} we study solutions of Bogomolny equations on 
$\RR^2\times \SS^1$ with prescribed singularities. Their moduli spaces provide 
more examples of novel self-dual gravitational instantons and are related to
four-dimensional ${\cal N}=2$ gauge theories with matter compactified on a circle.

In the remainder of this section we define periodic monopoles more precisely
and formulate our main result, the correspondence between periodic monopoles and solutions of Hitchin equations on a cylinder.

\subsection{Periodic Dirac Monopoles}

Before investigating the nonabelian Bogomolny equation on $\RR^2\times \SS^1$,
it is instructive to write down solutions of the Bogomolny equation
in the case $G=U(1)$. In this case there are no nontrivial smooth solutions, 
so we allow for singularities at some finite number of points on $\RR^2\times
\SS^1$. A solution with one 
singularity represents a Dirac monopole on $\RR^2\times \SS^1$. 

For $G=U(1)$ the Bogomolny
equation implies that the Higgs 
field satisfies the Laplace equation
$$
\nabla^2\phi=0.
$$
Let $z$ be a complex affine coordinate on $\CC\simeq \RR^2$ and $\chi\in [0,2\pi]$
be the periodic coordinate on $\SS^1$. We will denote by $x$ the pair $(z,\chi)$.
The solution corresponding to a Dirac monopole at $x=0$ is given by
$$
\phi(x)=v+kV(x)\equiv 
v+k\frac{\log(4\pi)-\gamma}{2\pi}-\frac{k}{2}{\sideset{}{'}\sum_{p=-\infty}^{\infty}} \left[ 
\frac{1}{\sqrt {|z|^2+(\chi-2\pi p)^2}}-\frac{1}{2\pi |p|}\right],
$$
where the prime means that for $p=0$ the second term in the square brackets must
be omitted, and $\gamma$ is the Euler's constant.
$V(x)$ satisfies the Laplace equation everywhere except $z=0, \chi=0 \mod 2\pi$.
Near this point $V(x)$ diverges:
$$
V(x)\sim -\frac{1}{2{\sqrt{|z|^2+|\chi|^2}}}+O(1).
$$
For large $|z|$ the function $V(x)$ is given by
$$
V(x)\sim \frac{\log |z|}{2\pi}+o(1).
$$
The connection $A$ corresponding to this Higgs field has the following asymptotics
for $|z|\ra\infty$ (up to a gauge transformation):
$$
A_z\sim \frac{a}{z}+o(1/z),\ A_\chi=\frac{k}{2\pi}\arg z + b+o(1).
$$
Here $a$ and $b$ are real constants. For these formulas to define a connection on a 
$U(1)$ bundle, the parameter $k$ must be an integer.

The magnetic charge of a $U(1)$ monopole is defined as the first Chern class of the
monopole bundle restricted to the 2-torus $|z|=R$ for sufficiently large $R$.
It is easy to see that the magnetic charge of the above monopole is $k$.

The Higgs field of a solution describing several periodic Dirac monopoles has the form
\begin{equation}
\phi(x)=v+\sum_\alpha k_\alpha V(x-x_\alpha).
\end{equation}
It is singular at $x=x_\alpha$ and for large $|z|$ behaves as
\begin{equation}
\phi(x)\sim v+\frac{\log |z|}{2\pi}{\sum_\alpha k_\alpha} +o(1).
\end{equation}

\subsection{Asymptotics Of A Periodic Monopole}

It is well known that finite-energy solutions of $SU(2)$ Bogomolny equations on 
$\RR^3$ are exponentially  close to the Dirac monopole at large distances~\cite{JT}. 
Then it is natural to require that periodic $SU(2)$ monopoles be close to the periodic 
Dirac monopole at large $|z|$. Accordingly, we will look for solutions of $SU(2)$ Bogomolny 
equations on $\RR^2\times \SS^1$ such that outside a compact set $T\subset 
\RR^2\times \SS^1$ one has
\begin{eqnarray}\label{asympt}
\phi(x) & \sim & g(x)\ \sigma_3\ \phi_D(x)\ g(x)^{-1}+o(1),\nonumber\\
d_A\phi(x) & \sim & g(x)\ \sigma_3\ d\phi_D(x)\ g(x)^{-1}+o(1/|z|),\nonumber\\
A(x)    & \sim & g(x)\ \sigma_3\ A_D(x)\ g(x)^{-1} + g(x) dg^{-1}(x)+o(1).
\end{eqnarray}
Here $\sigma_3=\diag \left\{1,-1\right\}$,\  $g(x)$ is an $SU(2)$-valued function on 
$\left(\RR^2\times\SS^1\right)\backslash T$, and $\phi_D$ and $A_D$ are a 0-form and
a $U(1)$ connection defined by
\begin{eqnarray}
\phi_D(x) &=& v+k\frac{\log |z|}{2\pi},\label{phiD} \\
A_D(x)   &=& b+\frac{k}{2\pi}\arg z.\label{AD}
\end{eqnarray}
This means that up to terms vanishing at infinity a periodic $SU(2)$ monopole 
is gauge-equivalent to a periodic Dirac monopole with charge $k$ embedded in a 
$U(1)$ subgroup of $SU(2)$. 

The real parameters $v$ and $b$ will often appear in a combination $v+ib$.
We will denote this combination $\ups.$

Note that we implicitly set the circumference of circle parameterized by $\chi$ to be
$2\pi$. This does not entail a loss of generality, as the Bogomolny equation
is invariant with respect to rescalings of the metric on $X$. One has to keep in mind
that rescaling the circumference by a factor $\lambda$ requires rescaling the
Higgs field $\phi$ by the same factor. Thus the ``large circumference limit''
is equivalent to the ``large $v$ limit.'' We will use these terms interchangeably.

The magnetic charge of a periodic monopole is defined in analogy
with the case of monopoles on $\RR^3$. It follows from Eq.~(\ref{asympt})
that for large enough $|z|$ the eigenvalues of $\phi$ are distinct (and opposite).
Hence for large enough $|z|$ one has a well-defined line bundle $L_+\subset E$, the
eigenbundle of $\phi$ associated with the positive eigenvalue. The magnetic
charge can be defined as the first Chern class of $L_+$ restricted to a 2-torus
$|z|=R$, where $R$ is large enough. Thus the magnetic charge
is given by the formula
\begin{equation}
m(A,\phi)=\lim_{R\ra\infty} \int_{|z|=R} \frac{\Tr\left( F_A\phi\right)}{4\pi||\phi||}.
\end{equation} 
It is easy to see that the magnetic charge of a periodic monopole is
nonnegative.
Substituting the asymptotics Eqs.~(\ref{asympt}-\ref{AD}) into this formula,
one finds that $m(A,\phi)=k$, so $k$ must be nonnegative too. 
Unlike the case of monopoles on $\RR^3$,
the energy of a monopole is infinite for $k\neq 0$.

\subsection{Nahm Transform For Periodic Monopoles}

Let $\Sigma$ be a Riemann surface, $V$ be a unitary vector bundle on $\Sigma$,
$\hat{A}$ be a connection on $V$, and $\Phi$ be a section of 
$\End (V)\ot \Omega_\Sigma^{1,0}.$ 
Hitchin equations for $\hat{A},\Phi$ are the equations
$$
\bpartial_{\hat{A}}\Phi=0,\qquad F_{\hat{A}}+\frac{i}{4}[\Phi,\Phi^\dag]=0,
$$
where the commutator is understood in a graded sense, i.e.
$[\Phi,\Phi^\dag]=\Phi\wedge \Phi^\dag+\Phi^\dag\wedge\Phi.$
Hitchin equations are the reduction of the self-duality equation to two
dimensions.

Our main result is that there is a one-to-one correspondence (modulo gauge 
transformations) between $SU(2)$ periodic monopoles with magnetic charge $k$
and solutions of $U(k)$ Hitchin equations on a cylinder $\RR\times \SS^1$
with the Higgs field growing exponentially at infinity. To describe the asymptotics
of the Higgs field more precisely, let us regard $\RR\times\SS^1$ as a strip
$0\leq {\rm Im}\ s\leq 1$ on the complex $s$--plane with the boundaries
identified in an obvious manner. Then the Higgs field behaves as follows
for ${\rm Re}\ s\ra\pm\infty$:
\begin{equation}\label{hasympt}
\Phi(s)\sim g_{\pm}(s,\bs)\ e^{\pm \frac{2\pi s}{k}}\  
\diag(1,\omega,\omega^2,\ldots,\omega^{k-1})\ g(s,\bs)_{\pm}^{-1}\ ds.
\end{equation}
Here $g(s,\bs)_\pm$ are some (multi-valued) functions with values in $U(k)$,
and $\omega$ is a $k$--th root of unity. The curvature of the $U(k)$ connection, 
on the other hand, approaches zero as $1/|{\rm Re\ s}|^{3/2}$ for 
${\rm Re\ s}\ra\pm\infty.$

For $k=1$ it is easy to write down an explicit solution of Hitchin equations
with this asymptotics. Then Nahm transform implies that there exists a
periodic $SU(2)$ monopole with $k=1$. One can also
argue that solutions of Hitchin equations exist for all positive $k$, and even describe
their moduli space. This implies that periodic monopoles exist for all $k>0$.
It would be interesting to find an explicit formula for periodic monopole,
at least in the $k=1$ case.

\subsection{Outline}

The paper is organized as follows. In Section~\ref{sec:gaugetheory} we
explain the relation between periodic monopoles and ${\cal N}=2$ super Yang-Mills
theory compactified on a circle. This section requires familiarity
with the physics of branes in Type II string theory. The rest of the paper
does not depend on it.

In Section~\ref{sec:Nahm} we show that the Nahm transform takes
periodic monopoles to solutions of Hitchin equations on a cylinder.
In Section~\ref{sec:specS} we explain how to associate algebro-geometric
data to a periodic monopole. These data consist of an algebraic curve
and a line bundle over it and are important in the study of
Nahm transform. On the other hand, it is well known that
to every solution of the Hitchin equations one can associate
so-called spectral data also consisting of an algebraic curve
and a line bundle. In Section~\ref{sec:CandS} we show that the 
algebro-geometric data associated to the periodic monopole
coincide with the spectral data of its Nahm transform.
In Section~\ref{sec:asymptotics} we use this information to determine
the asymptotic behavior of the solutions of Hitchin equations
arising from periodic monopoles. 
In Section~\ref{sec:Inverse} we describe the ``inverse'' Nahm
transform which produces a solution of the Bogomolny equation on $\RR^2\times\SS^1$
from a solution
of Hitchin equations on a cylinder. In Section~\ref{sec:invspec} we study the
asymptotic behavior of the resulting solution of the Bogomolny equation and show that
it is given by~(\ref{asympt}). In Section~\ref{sec:Nahmsq}
we prove that the composition of the ``direct'' and ``inverse''
Nahm transform takes a periodic monopole to a gauge-equivalent
periodic monopole. The proof is modelled on that of Schenk~\cite{Schenk} and 
requires rather tedious computations. Another approach to the proof which uses the
spectral sequence technology is sketched in the Appendix.
The results of Sections~\ref{sec:Nahm}-\ref{sec:Nahmsq} imply that the Nahm transform establishes
a one-to-one correspondence between periodic monopoles and 
solutions of Hitchin equations on a cylinder with a particular
asymptotics. In Section~\ref{sec:existence} we give (nonrigorous)
arguments that periodic monopoles exist for all $k>0$ and are (almost) completely
determined by their spectral data.
Assuming that this is true, we show in Section~\ref{sec:modulispace} that
the centered moduli space of a charge $k$ periodic monopole has dimension
$4(k-1)$, and describe a distinguished complex structure on it. We also
argue that the centered moduli space carries a natural hyperk\"ahler metric.

\section{Periodic Monopoles And Brane Configurations}\label{sec:gaugetheory}

This section assumes familiarity with the Chalmers-Hanany-Witten-type brane 
configurations~\cite{ChH,HW,Witten} and their use in solving quantum gauge theories with eight supercharges.

Consider two parallel flat NS5-branes in Type IIB string theory. 
For definiteness, let us assume that their worldvolumes are given by
the equations 
$$
x^6=x^7=x^8=x^9=0
$$
and 
$$
x^6=v,\ x^7=x^8=x^9=0.
$$
This brane configuration is BPS (preserves sixteen supercharges), and
its low-energy dynamics is described by a $d=6$ supersymmetric 
Yang-Mills theory with gauge group $U(2)$. 

Consider now a D3-brane with the worldvolume given by 
$$
x^3=x^4=x^5=x^7=x^8=x^9=0,\ 0\leq x^6\leq v.
$$ 
This is an open D3-brane, in the sense
that its worldvolume has boundaries. This is possible because the boundaries
lie on the NS5-branes. One can say that such a D3-brane is suspended
between the NS5-branes.

{}From the point of view of the Yang-Mills theory describing the NS5 branes,
the suspended D3-brane is a static solution of the Yang-Mills equations
of motion with a unit magnetic charge~\cite{ChH}.
Moreover, since a suspended D3-brane preserves eight supercharges, it is a BPS soliton, and must solve the Bogomolny equation. 

Similarly, $k$ suspended D3-branes are described in the Yang-Mills theory
by a charge $k$ monopole~\cite{ChH}. 

Let us now compactify the $x^3$ coordinate on a circle of radius
$R$, i.e. let $x^3$ take values in $\RR/(2\pi R\cdot\ZZ)$ rather than in $\RR$.
In such a situation
we may still consider suspended D3-branes. The same arguments as in the
uncompactified case lead one to the conclusion that $k$ D3 branes
are described in the Yang-Mills theory by a BPS monopole on $\RR^2
\times \SS^1$ with charge $k$. The $\SS^1$ has circumference 
$2\pi R$.

Now let us apply T-duality in the $x^3$ direction. This
has the effect of taking us to Type IIA string theory. We will denote
the spatial coordinates in Type IIA by $y^1,\ldots y^9$, so that
$y^3$ can be identified with the Fourier dual of $x^3$, while the rest
of the $y$ coordinates are identified with the corresponding
$x$ coordinates. If we choose the units in which the Regge slope
$\alpha'$ is unity, then $y^3$ has period $2\pi/R$.

The usual T-duality rules tell us that the Type IIB NS5-branes
are mapped under T-duality to the Type IIA NS5 branes with the worldvolumes
given by
$$
y^6=y^7=y^8=y^9=0
$$
and
$$
y^6=v,\ y^7=y^8=y^9=0.
$$
A suspended D3-brane is mapped to a D4-brane with the worldvolume given by
$$
y^4=y^5=y^7=y^8=y^9=0,\ 0\leq y^6\leq v.
$$
This D4-brane is suspended between the NS5-branes. 

Such a brane configuration in Type IIA string theory has been first
studied by E.~Witten~\cite{Witten} and subsequently by many other authors. 
The only difference with~\cite{Witten}
is that in our case $y^3$ is a periodic variable.
Witten argued that the low-energy dynamics of $k$ suspended D4-branes
is described by the $d=4$ ${\cal N}=2$ supersymmetric Yang-Mills theory
with gauge group $SU(k)$. The classical gauge coupling of this theory
depends on $v$: $1/g_{YM}^2\sim v/g_{st}$. Thus we are dealing with a four-dimensional
super Yang-Mills theory on $\RR^3\times \SS^1$ where the circumference
of $\SS^1$ is given by $2\pi/R$.

In the quantum theory the gauge coupling depends on the renormalization
scale $\mu$: $1/g_{YM}^2(\mu)=1/g_{YM}^2(\mu_0)+k\log(\mu/\mu_0)$.
{}From the string theory viewpoint, taking into account quantum corrections
on the D4-brane worldvolume is equivalent to taking into account
the back-reaction of the D4-branes on the NS5-branes. This back-reaction
results in the bending of the NS5-branes, as a consequence of which
the distance between them in the $y^6$ direction starts to depend on 
$u=y^4+iy^5$:
$$
\delta x^6(u)=const+k\ {\rm Re}\log u.
$$ 

Since D3-branes suspended between NS5-branes in Type IIB string
theory are T-dual to D4-branes suspended between NS5-branes in
Type IIA string theory, their moduli spaces must coincide (as
Riemann manifolds).
The moduli space of the former coincides with the moduli space
of $k$ periodic monopoles. The moduli space of the latter
is the Coulomb branch of the $d=4$ ${\cal N}=2$ supersymmetric Yang-Mills
with gauge group $SU(k)$ compactified on a circle of radius $1/R$.

Assuming that this correspondence is true, we may predict the
dimension of the moduli space of periodic monopoles. As explained
in~\cite{SWthree}, the Coulomb branch of a $d=4$ ${\cal N}=2$ super-Yang-Mills
theory on $\RR^3\times \SS^1$ is a 
hyperk\"ahler manifold of dimension
$4\,{\rm rank}(G)$, where $G$ is the gauge group. Thus the moduli space of a periodic 
monopole of charge $k$ must have real dimension $4(k-1)$.

A particular case of this correspondence has been known for some
time from the work of Chalmers and Hanany~\cite{ChH}. These authors
showed that the centered moduli space of $k$ periodic monopoles
on $\RR^3$ coincides with the Coulomb branch of the
$d=3$ ${\cal N}=4$ supersymmetric Yang-Mills with gauge group $SU(k)$
on $\RR^3$. This statement follows from ours in the limit $R\ra \infty$.
In this limit monopoles on $\RR^2\times\SS^1$ reduce to ordinary monopoles
on $\RR^3$. On the other hand, the radius of the dual circle goes
to zero, and therefore the $d=4$ ${\cal N}=2$ super-Yang-Mills undergoes Kaluza-Klein
reduction to the $d=3$ ${\cal N}=4$ super-Yang-Mills.

We pause here to explain one subtlety in the above arguments. The Coulomb branch 
of the $d=3$ super-Yang-Mills theory with gauge group $SU(k)$ is 
related to the centered monopole moduli space \cite{ChH, HW}, while the Coulomb branch
of the $d=4$ gauge theory on a circle appears to be related to the uncentered moduli space of periodic monopoles. If this were the case, we would not
get an exact agreement between the two statements in the limit $R\ra \infty$.
In fact, when considering periodic monopoles, one is forced to fix their
center-of-mass if one wants to get a well-defined metric on the
moduli space. The reason is that the translational zero modes of a
periodic monopole are not normalizable. This is explained in more detail
in Section~\ref{sec:modulispace}. 
In this way the contradiction is avoided.
(The fact that the translational zero modes
for suspended D4 branes are not normalizable was explained from the
string theory point of view in~\cite{Witten}. This ''freezing out''
of the center-of-mass motion is the ultimate reason why the suspended
D4-branes are described by an $SU(k)$ rather than $U(k)$ gauge theory.)

Another interesting limit is $R\ra 0$. In this limit the circle on
which the $d=4$ ${\cal N}=2$ super-Yang-Mills theory is compactified
becomes arbitrarily large, while the monopole interpretation looses
meaning. The Coulomb branch of this theory with all quantum corrections  
has been determined in~\cite{SW,ArF}.
It is a special K\"ahler manifold of real dimension $2(k-1)$. (Note
that the dimension of the Coulomb branch jumps by a factor two
as soon as one compactifies one dimension a circle. The reason for this is explained
in~\cite{SWthree}.) The simplest way to derive the answer uses the
Type IIA brane configuration with suspended D4-branes described 
above~\cite{Witten}. One notices that the metric 
on the Coulomb branch does not depend on the string coupling if $g_{YM}$ is kept
fixed, so
one can consider the limit $g_{st}\ra \infty, v\ra \infty.$ In this limit Type
IIA string theory reduces to $d=11$ supergravity, and the configuration
with D4-branes suspended between two NS5 branes turns into a
single smooth M5-brane. The metric on the Coulomb branch with all quantum
corrections taken into account can be obtained by a classical
computation with an M5-brane.

It would certainly be nice if the quantum Coulomb branch of the compactified
theory could also be determined by a classical computation in
$d=11$ supergravity. However, it is easy to see that this is not the case.
The reason is that upon compactification on a circle there appear
new kinds of instantons in the gauge theory, namely virtual BPS
monopoles and dyons whose worldlines wrap the compactified circle.
In a strongly coupled string theory such effects are captured by
membrane instantons. These instantons are represented by
Euclidean open M2-branes whose boundaries lie on the M5-brane.
Clearly, directly summing up all such instantons is a hopeless
task.

Nevertheless, one can give a ``classical'' 
recipe for computing the complete quantum Coulomb branch of the compactified
super-Yang-Mills theory by exploiting the correspondence with periodic
monopoles. Computing the metric on the moduli space of periodic monopoles
is a well-defined problem which appears much simpler than summing up membrane instantons. One could hope to determine this metric using twistor
methods, similarly to how it has been done for ordinary monopoles.
Alternatively, one could apply Nahm transform to make the problem
more manageable. Below we show that the Nahm transform of a periodic
monopole is described by Hitchin equations on $\CC^*$. These equations
are somewhat simpler than the original Bogomolny equation. The
properties of the moduli space of periodic monopoles will be studied
in detail in a forthcoming publication~\cite{usthree}.

\section{From Periodic Monopoles To Solutions Of Hitchin Equations}\label{sec:Nahm}

In this section we show that Nahm transform associates to every periodic $SU(2)$ 
monopole with charge $k$ a solution of $U(k)$ Hitchin equations on a cylinder.
We follow~\cite{BraamBaal} where the Nahm transform for instantons
on $T^4$ is discussed. In fact, periodic monopoles can be regarded as a 
limiting case of instantons on $T^4$ invariant with respect to a subgroup of translations.
Another closely related work is~\cite{Marcos}, where Nahm transform for instantons
on $\RR^2\times T^2$ is studied. We will use many of the techniques of~\cite{Marcos} 
and~\cite{DK}.

Let the pair $(A,\phi)$ be a periodic $SU(2)$ monopole with 
asymptotics~(\ref{asympt}). Let $\S$ be the spinor bundle on 
$X=\RR^2\times \SS^1$. This means that $\S$ is a trivial unitary rank 2 bundle on $X$
equipped with an injective bundle morphism $\sigma: T^*X\to S\ot S^*$ which is 
Hermitian and has zero trace.
By a change of trivialization, one can always bring $\sigma$ to the standard form
$\sigma(dx_j)=\sigma_j,\ j=1,2,3,$ where $\sigma_j$ are the Pauli matrices.
Let $L$ be a trivial unitary line bundle on $X$ with a flat unitary connection $a$
whose monodromy around $\SS^1$ is $\exp(-2\pi it),\ t\in \RR/\ZZ$ 
(these conditions define a unique connection).

Consider a Dirac--type operator $D:E\ot S\ot L \to E\ot S\ot L$ of the form
\begin{equation}
D=\sigma\cdot d_{A+a}-(\phi-r).
\end{equation}
We will be interested in its $L^2$ kernel and cokernel.
Using the fact that the norm of the Higgs field $\phi$ grows logarithmically at infinity,
one can show that $D$ is Fredholm for any $(r,t)\in\RR\times \RR/\ZZ$. 
Thus its 
$L^2$--index is independent of $r,t$. As explained in the end of this section, 
the index is equal to the negative of the magnetic charge $k$.

The Weitzenbock formula for $D$ reads:
\begin{equation}\label{Wone}
D^\dagger D=-\nabla_{A+a}^2+(\phi-r)^2+\sigma\cdot(d_A\phi-*F_A).
\end{equation}
This formula together with the Bogomolny equation imply that $D^\dagger D$ is
a positive-definite operator, and therefore $D$ has a trivial $L^2$ kernel.
It follows that $\Ker\ D^\dagger$ is a rank $k$ trivial bundle over the 
$(r,t)$--plane. Actually, since $t$ is a periodic variable, we get a rank 
$k$ bundle over a cylinder $\hX=\RR\times \SS^1\cong\CC^*.$ This trivial
bundle will be denoted $\hE$.

{}From the growth of $\phi$ at infinity it follows that for all $s=r+it$ the elements 
of $\Ker\ D^\dag$ decay at least exponentially. Thus for all $s\in\CC$ we have a 
well-defined Hermitian inner product on $\hE_s$.  If we choose a basis 
$\psi_1(x,s),\ldots,\psi_k(x,s),\ x\in X,$ of
$\Ker\ D^\dag$ at point $s$, then the explicit formula for the inner product is
$$
\lg \psi_\alpha,\psi_\beta\rg=\int \psi_\alpha(x,s)^\dag \psi_\beta(x,s) d^3x.
$$
This inner product makes $\hE$ into a unitary bundle.
Below it will be assumed that the vectors $\psi_\alpha,\,\alpha=1,\ldots,k,$ are chosen to 
form an orthonormal
basis of $\Ker\ D^\dag$ for all $s$.

Next we want to define
a connection $\hA$ on $\hE$ and a Higgs field $\hphi\in \Gamma(\End(\hE))$.
The Higgs field at a point $s\in \hX$ is a linear map from $\hE_s$ to $\hE_s$.
We define this map as a composition of two maps: multiplication by
$z$ and projection to $\hE_s.$ 
An explicit formula for $\hphi$ in an orthonormal basis is
\begin{equation}
\hphi(s)^\alpha_\beta=\int \psi_\alpha(x,s)^\dagger\ z\ \psi_\beta(x,s) d^3x. 
\end{equation}
Since all $\psi_\alpha$ decay at infinity faster than any power of $z,$
this is well-defined.

The connection $\hA$ on $\hE$ is induced by the zero connection on the trivial
infinite-dimensional bundle whose fiber at a point $s\in \hX$ consists of
all smooth $L^2$ sections of $E\ot S\ot L$. In components:
\begin{equation}
\hA_s(s)^\alpha_\beta ds=i\int \psi_\alpha(x,s)^\dagger\ 
ds\left(\frac{\partial}{\partial s}\psi_\beta(x,s)\right)\ d^3x. 
\end{equation}

It is easy to see that $\hA$ is a unitary connection on $\hE$. 
As for $\hphi\in \Gamma(\End(\hE))$, it is not Hermitian, unlike its counterpart 
$\phi\in \Gamma(\End(E))$.

Now we will show that $\hA$ and $\hphi$ satisfy Hitchin equations.
We will need the following commutation relations:
\begin{equation}\label{commrel}
\begin{array}{llll}
{}[D,z]=\sigma_+, & [D,\bz]=\sigma_-, & [D,\partial]=-p_+,
& [D,\bpartial]=-p_-, \\
{}[D^\dag,z]=-\sigma_+, & [D^\dag,\bz]=-\sigma_-, & [D^\dag,\partial]=-p_-, &
[D^\dag,\bpartial]=-p_+.
\end{array}
\end{equation}
Here $\partial=\partial/\partial s, \bpartial=\partial/\partial \bs,$
$\sigma_\pm=\sigma_1\pm i\sigma_2,$ $p_\pm=\frac12 (1\pm \sigma_3)$.
We will denote the projector to $\Ker\ D^\dag$ by $P$. Its explicit form is
$$P=1-D(D^\dag D)^{-1} D^\dag.$$ The projector to the orthogonal complement
of $\Ker\ D^\dag$ will be denoted by $Q$:
$$Q=D(D^\dag D)^{-1} D^\dag.$$ 

{}First let us compute $\bpartial_\hA \hphi$:
$$
\begin{array}{lcl}
\bpartial_\hA \hphi & = & [P\bpartial,Pz] \\
& = & (P\bpartial (1-Q) z-Pz(1-Q)\bpartial) \\
& = & P(zQ\bpartial-\bpartial Q z).
\end{array}
$$
Using the identity
$PD=0,$ and keeping in mind that $\bpartial_\hA \hphi$\ should be thought of
as acting on $\Ker\ D^\dag$ from the right, we can rewrite this expression as follows:
$$
\begin{array}{lcl}
\bpartial_\hA \hphi & = & P([z,D](D^\dag D)^{-1} [D^\dag,\bpartial]-
[\bpartial, D] (D^\dag D)^{-1} [D^\dag, z])\\
& = &  P(\sigma_+ (D^\dag D)^{-1}p_+ + p_-(D^\dag D)^{-1} \sigma_+).
\end{array}
$$
To go from the first line to the second line we used the commutation relations
(\ref{commrel}). 
The Weitzenbock formula~(\ref{Wone}) tells us that $D^\dag D$ commutes with 
all $\sigma_j,\ j=1,2,3$, and since $p_-\sigma_+=\sigma_+ p_+=0$, we get the ``complex'' Hitchin equation
\begin{equation}\label{Hitcomp}
\bpartial_{\hA}\hphi=0.
\end{equation}

The curvature $\hF$ of the connection $\hA$ is given by
$$
\hF=i[P\partial,P\bpartial] ds\wedge d\bs.
$$
We can simplify this as 
follows:
\begin{equation}\label{FHitchin}
\begin{array}{rcl}
\hF & = & i P(\partial Q\bpartial Q-\bpartial Q\partial Q)ds\wedge d\bs \\
& = & i P(\partial D (D^\dag D)^{-1}\bpartial D^\dag-
\bpartial D (D^\dag D)^{-1}\partial D^\dag)ds\wedge d\bs \\
& = & i P (p_+(D^\dag D)^{-1}p_+ - p_-(D^\dag D)^{-1} p_-)ds\wedge d\bs \\
&= & i P (D^\dag D)^{-1} \sigma_3 ds\wedge d\bs.
\end{array}
\end{equation}
Here we again used the commutation relations~(\ref{commrel}) and the fact that
$D^\dag D$ commutes with all $\sigma_j$. On the other hand, let us compute the
commutator $[\hphi,\hphi^\dag]$:
$$
\begin{array}{lcl}
{}[\hphi,\hphi^\dag] & = & [Pz,P\bz] \\
{}& = & (P\bz Q z-PzQ\bz) \\
{}& = & P([\bz,D](D^\dag D)^{-1}[D^\dag,z]-[z,D](D^\dag D)^{-1}[D^\dag,\bz])\\
{}& = & P(\sigma_- (D^\dag D)^{-1}\sigma_+-\sigma_+(D^\dag D)^{-1}\sigma_-)\\
{}& = & -4P(D^\dag D)^{-1}\sigma_3.
\end{array}
$$
Comparing with the expression for the curvature of $\hA$, we obtain the
``real'' Hitchin equation:
\begin{equation}\label{Hitreal}
\hF_{s\bs}+\frac{i}{4}[\hphi,\hphi^\dag]=0.
\end{equation}
To make the last equation covariant with respect to diffeomorphisms of 
$\hX$ one should think of $\Phi=\hphi\ ds$ as a section of $\End(\hE)\ot 
\Omega^{1,0}_\hX$.
Then the ``real'' Hitchin equation takes the form
\begin{equation}\label{Hitrealcov}
\hF+\frac{i}{4}[\Phi,\Phi^\dagger]=0,
\end{equation}
where the commutator is understood in the graded sense.

{}Following~\cite{HitchinSpec}, we can associate to any solution of Hitchin equations
an algebraic curve $\C.$ In the present case the curve is a hypersurface
in $\CC\times\CC^*$ defined by the equation 
\begin{equation}\label{C}
\det(z-\hphi(s))=0.
\end{equation}
Here $z$ is an affine parameter on $\CC$, while $s$ parameterizes $\hX\cong\CC^*.$
The left-hand-side of the above equation is a polynomial in $z$ of degree $k$,
and it follows from the ``complex'' Hitchin equation that its coefficients
are holomorphic functions on $\CC^*$. This shows that the above equation
defines an algebraic curve which is noncompact and is a $k$--fold cover of $\CC^*$.

The eigenvectors of $\hphi$ obviously form a sheaf $N$ on $\C$ whose stalk at a general point
is one-dimensional. The direct image of $N$ under the projection map $\pi:\C\to \CC^*$ is 
the bundle $\hE$. We will call the pair $(\C,N)$ the spectral data of a Hitchin pair $(\hA,\hphi)$,
and refer to $\C$ as the Hitchin spectral curve.

{}For a general Hitchin pair the curve $\C$ is nonsingular. If this is the case, then the sheaf $N$ is 
a line bundle. Indeed, since $\pi_*(N)$ is a vector bundle, $N$ is a torsion free sheaf,
hence a subsheaf of a locally free sheaf. But any subsheaf of a locally free sheaf on a smooth 
algebraic curve is locally free (this follows from the fact that a nonsingular curve has 
cohomological dimension one). Thus $N$ must be a line bundle.

{}Finally, let us justify the assertion that $\ind\ D=-k$. The index can be computed 
using the heat kernel method. Alternatively one may use the approach of Callias~\cite{Callias} who
computed the index of a Dirac-type operator on $\RR^{2n+1}$ for all $n$. One can check
that the proof goes through for $\RR^2\times\SS^1$. Either way, we find:
$$
\begin{array}{lcl}
\ind\ D &=& \displaystyle{\lim_{R\ra\infty}-\int_{|z|=R}
\frac{\Tr\ (*(\partial_A\phi)\phi)}{4\pi||\phi||}}\\
&=&\displaystyle{\lim_{R\ra\infty} -\frac{1}{2\pi} \int_{|z|=R} 
\frac{\partial}{\partial r}||\phi||\ d(\arg z)\wedge d\chi.}
\end{array}
$$
Thus $\ind\ D=-m(A,\phi)=-k.$ Below we will compute the index in another way, which also
provides some information on the spatial structure of the zero modes.

\section{Spectral Data Of A Periodic Monopole}\label{sec:specS}

In the previous section we showed that the Nahm transform of a charge $k$ periodic 
monopole is a pair $(\hA,\Phi)$, where $\hA$ is a connection on a trivial rank $k$ 
bundle $\hE$ over $\hX=\RR\times \SS^1,$ $\Phi$ is a section of 
$\End (\hE)\ot \Omega^{1,0}_\hX$, and the pair $\hA,\Phi$ satisfies the Hitchin 
equations~(\ref{Hitcomp},\ref{Hitreal}). Since $\hX$ is noncompact, it is
important to determine the behavior of the pair $(\hA,\Phi)$ at $r=\pm\infty$.
The simplest way to do this uses an algebraic curve associated to the periodic 
monopole. In this section we explain how to construct this curve and
a line bundle over it. These algebro-geometric data associated to
a periodic monopole will be called the monopole spectral data.

Let $B$ be a (nonunitary) 
connection on $E$ defined by
$$
B(x)=A(x)-i\phi(x) d\chi.
$$
Let $\zeta\in \CC$. Consider a loop $\gamma_{\zeta}: \SS^1\to X$ given by 
$$\gamma_{\zeta}: u\ra (z(u),\chi(u))=(\zeta,u),\qquad u\in \RR/2\pi\ZZ.$$ 
We denote the value of $B(\gamma_{\zeta}(u))$ on the vector $\partial/\partial u$ by 
$B_u$. Suppose we want
to compute the holonomy of $B$ along $\gamma$. To do this, we must solve the
matrix equation
\begin{equation}\label{holonomy}
\left(\frac{d}{du}-iB_u\right)V(\zeta,u)=0
\end{equation}
with the initial condition $V(\zeta,0)=1_{2\times 2}.$ The holonomy is equal to
$V(\zeta,2\pi).$ 

Note now that the Bogomolny equation implies
$$
\left[\partial_{\bz}-iA_{\bz}(\zeta,u),\frac{d}{du}-iB_u\right] =
-iF_{\bz\chi}-(\partial_{\bz}\phi-i[A_{\bz},\phi])\vert_{z=\zeta}=0.
$$ 
Hence the commutator 
$$
W(\zeta,u)=\left[\partial_{\bz}-iA_{\bz},V(z,u)\right]\vert_{z=\zeta}
$$
also satisfies the differential equation (\ref{holonomy}).
On the other hand, since $V(\zeta,0)=1_{2\times 2}$ for all $\zeta$,
$W(\zeta,0)=0$ for all $\zeta$. The equation (\ref{holonomy}) being first order,
this means that $W(\zeta,u)=0$ for all $\zeta\in \CC$ and $u\in \RR$. Recalling the
definition of $W$, we see that the characteristic polynomial of $V(z,u)$
is a holomorphic function of $z$ for any $u$. Hence the function
$$
F(w,z)=\det(w-V(z,2\pi))
$$
is a holomorphic function of both $z$ and $w$. It is also easy to see that $F(w,z)$ is
gauge-invariant and independent of the choice of origin on the circle
parameterized by $\chi$. 

We define the spectral curve $\Sc$ of a periodic monopole
to be the zero set of $F(w,z)$, i.e. $\Sc$ is an algebraic curve in $\CC^2$
given by
\begin{equation}\label{moncurve}
\det(w-V(z,2\pi))=0.
\end{equation}
Since both $\phi$ and $A$ are traceless, $\det V(z,2\pi)=1$. It follows that
$\Sc$ does not have common points with the set $w=0$ in $\CC^2$, and therefore may be 
regarded as a complete curve in $\CC\times \CC^*,$ where $\CC^*$ is a complex $w$--plane with the origin removed.

Let us examine the curve $\Sc$ more closely. Since we are dealing with $SU(2)$ monopoles,
the equation of $\Sc$ is really
\begin{equation}\label{curve}
w^2-w\ \Tr\ V(z,2\pi)+1=0,
\end{equation}
i.e. $\Sc$ is a double cover of the $z$--plane. One can also show that $\Tr\ V(z,2\pi)$
is a degree $k$ polynomial in $z$. Indeed, we already know that $\Tr\ V(z,2\pi)$ is
an entire function of $z$. Its behavior for large $z$ can be computed from the known
behavior of $A$ and $\phi$ described by (\ref{asympt}). This yields
\begin{equation}
\Tr\ V(z,2\pi)=z^k\ \exp\left(2\pi\ups\right)\ (1+o(1)),
\end{equation}
with $\ups=v+i b$.
Since the function $\Tr\ V(z,2\pi)$ is entire and bounded by a multiple
of $z^k$, it must be a polynomial of degree $k$. The leading coefficient
of this polynomial is determined by the asymptotic conditions imposed
on the monopole (i.e. by $b$ and $v$), while the remaining $k$ coefficients
are the moduli of the periodic monopole.

A periodic monopole also provides us with a coherent sheaf $M$
on $\Sc$, namely the sheaf of eigenvectors of $V(z,2\pi)$. The stalk of $M$ at a general 
point is one-dimensional. The direct image of $M$ under the projection map 
$\pi:\Sc\to \CC$ is of course the bundle $E$ restricted to $\chi=0$.
We will call the pair $(\Sc,M)$ the spectral data of a periodic monopole.

{}For a general monopole the curve $\Sc$ is nonsingular. If this is the case, then $M$ is 
a line bundle. The reasoning leading to this conclusion is the same as for 
the Hitchin spectral data.

A periodic monopole with charge $k$ can be thought of as consisting
of $k$ monopoles of charge $1$. With the help of the spectral curve one may suggest
a precise definition of the location of these constituent monopoles on $\CC$.
These are the points where the holonomy $V(z,2\pi)$ has an eigenvalue $1$, i.e.
the roots of the equation $\Tr\ V(z,2\pi)=2.$ Since $\Tr\ V(z,2\pi)$ is a polynomial of degree
$k$, for a generic monopole this equation has $k$ distinct roots 
$\zeta_1,\ldots,\zeta_k$. We expect that when these points are well-separated, the
energy density is concentrated in their neighborhood. 

If we assume that the curve $\Sc$ is nonsingular, then at $z=\zeta_\alpha$ the
Jordan normal form of $V(z,2\pi)$ is
$$
\begin{pmatrix}
1 & 1 \\
0 & 1 
\end{pmatrix}.
$$
This implies that at $z=\zeta_\alpha$ the holonomy
$V(z,2\pi)$ has a single eigenvector with eigenvalue one. In other words, if we consider
the restriction of $E$ to the $\SS^1$ given by $z=\zeta_\alpha$, and equip it with a 
(nonunitary)
connection $B$, then this bundle has a covariantly constant section unique up to a scalar
multiplication. On the other hand, for other values of $\zeta$ the holonomy of $B$ has
both eigenvalues distinct from $1$, and the restriction of $E$ to the circle does not
have sections covariantly constant with respect to $B$. This elementary
observation plays an important role in the next section.

\section{Coincidence Of The Spectral Data}\label{sec:CandS}

The purpose of this section is to demonstrate that the two kinds of spectral data defined in sections \ref{sec:Nahm} 
and \ref{sec:specS} coincide. Recall that starting from
a periodic monopole $(A,\phi)$ twisted by 
$s=r+it\in \CC$ we defined a unitary bundle $\hE$ on $\CC^*$, formed 
by zero-modes of the twisted Dirac operator $D^\dag$, as well as a unitary connection on $\hE$, and a Higgs field 
$\hphi\in\Gamma(\End (\hE)).$ The Higgs field $\hphi(s)$ was defined as a composition of multiplication
by the affine coordinate $z$ on $\CC\cong\RR^2$ and projection to
$\Ker(D^\dag)$.
The coincidence of the spectral curves $\C$ and $\Sc$ is equivalent to the following 
statement: if $\zeta$ is an 
eigenvalue of the transformed Higgs field $\hphi$ at a point $s=\sigma$, then
$e^{2\pi\sigma}$ is an eigenvalue of the holonomy of $B=A-i\phi\ d\chi$ around the 
loop $\gamma_\zeta$ which winds around the $\SS^1$ at $z=\zeta$. This is
the statement that will be proved below. We will also show
that the zero modes of the Dirac operator $D^\dag$ are in one-to-one correspondence with the
points $\zeta\in \CC$ such that the restriction of $E$ to the circle $z=\zeta$ has a
covariantly constant section (with respect to the connection $B$). As explained in the previous section,
for a general monopole there are $k$ such points, so we see again that
$\dim\,\ker\,D^\dag=k$.

\subsection{Cohomological Description Of The Nahm Transform}\label{subsec:Coh}

We proceed to reformulate the Nahm transform of section \ref{sec:Nahm} in cohomological terms.
The benefits of such a reformulation will become apparent shortly. 
In particular the cohomological definition of the transformed Higgs field $\hphi$
is extremely simple.

Let us denote by $\Lambda^{0,1}(X,E)$ the bundles on $X=\RR^2\times\SS^1$ whose sections have the form
$fd\bz+g d\chi$, where $f,g\in \Gamma(E).$  $\Lambda^{0,2}(X,E)$ will denote the bundle
whose sections have the form $f d\bz\wedge d\chi$, where $f\in \Gamma(E)$.
The bundles $\Lambda^{0,1}(X,E)$ and
$\Lambda^{0,2}(X,E)$ are subbundles of the bundles of $E$-valued differential forms $\Lambda^1(X,E)$ 
and $\Lambda^2(X,E),$ respectively. Their names betray their origin in the Hodge
decomposition of forms on $\CC^2$. For uniformity of notation, we also set 
$\Lambda^{0,0}(X,E)=\Gamma(E)$.

Pursuing this analogy, we can identify  spinor bundles  $S^{+}(E)$ and $S^{-}(E)$
as follows:
\begin{equation}\label{decomposition}
S^{+}(E)=\Lambda^{0,0}(X,E) \oplus \Lambda^{0,2}(X,E),\ \ S^{-}(E)=\Lambda^{0,1}(X,E).
\end{equation}

To any trivial vector bundle $E$ on $X$ we can associate a locally free sheaf of vector 
spaces defined in the following way. Over the whole $X$ its space of sections is the space 
of smooth global sections of $E$ which belong to the Schwarz space (i.e. all of their
derivatives decay faster than any negative power of $|z|$). Over any open set 
${\mathcal O}\subset X$ its space of 
sections is obtained by restriction from $X$. In what follows we will identify a trivial 
vector bundle on $X$ and the corresponding sheaf.

Let us define differentials
\begin{equation}\label{Der}
\bD_p=\dbz\wedge 2 (\frac{\partial}{\partial{\bar{z}}}-iA_{\bz})+d\chi\wedge
(\frac{\partial}{\partial{\chi}}-iA_{\chi} -\phi+s),
\end{equation}
acting from $\Lambda^{0,p}(X,E)$ to $\Lambda^{0,p+1}(X,E)$, $p=0,1.$
Note that $\bD_1\bD_0=0$, as a consequence of the Bogomolny equation. Thus we have a differential 
complex $K$:
\begin{equation}\label{complex}
K:\ 0\ra \Lambda^{0,0}\xrightarrow{\bD_0}\Lambda^{0,1}\xrightarrow{\bD_1}
\Lambda^{0,2}\ra 0.
\end{equation}
Since the operators $\bD_p$ depend on $e^{2\pi s}\in \CC^*$, so does the complex $K$, and it would
be more precise to call it $K_s$. We will omit the subscript $s$ where this cannot lead
to confusion.

Since all the bundles we are dealing with are trivial, we are free to identify 
$S\ot E\cong S^{+}(E)\cong S^{-}(E)$. Then the twisted Dirac operator $D:S\ot E\ra S\ot E$
becomes simply
$$
D=\bD_0-\bD_1^*,
$$
and its adjoint $D^\dag: S^{-}\rightarrow S^{+}$ is $\bD_0^*-\bD_1$.

As explained in section~\ref{sec:Nahm}, the only $L^2$ solution of
the equation $D\psi=0,\ \psi\in \Gamma(S\otimes E),$ is the 
trivial one. In other words the equation $\bD_0\Psi-\bD_1^*\Psi=0$,
$\Psi\in\Lambda^{0,0}(X,E)\oplus\Lambda^{0,2}(X,E),$ has only the trivial
$L^2$ solution. It follows that the complex~(\ref{complex}) is exact in the first and the third terms: $H^0(K)=H^2(K)=0$. We want to show that $H^1(K)$ is isomorphic to the kernel of the twisted Dirac operator $D^\dag$.
In one direction this is easy: for any $\psi\in \Ker\ D^\dag$ we have $\bD_1\psi=\bD_0^*\psi=0$,
and therefore $\psi$ is a harmonic representative of a class in $H^1(K)$. 
It is obvious that this map from $\Ker\ D^\dag$ to $H^1(K)$ is injective. 
The inverse map is constructed as follows.
{}For any representative $\theta$ of a class $[\theta]\in H^1(K)$ we have to find $\rho\in \Lambda^{0,0}(X,E)$ such that
$(\bD_0^*-\bD_1)(\theta+\bD_0\rho)=0$. Since $H^0(K)$ is trivial, the kernel of the operator 
$\bD_0^*\bD_0$ is empty and the operator itself is invertible. Thus we may solve for the function $\rho$:
\begin{equation}
\rho=-(\bD_0^*\bD_0)^{-1}\bD_0^*\theta.
\end{equation}
This yields a map from $H^1(K)$ to $\Ker\ D^\dag.$ It is easy to see that it is the
inverse of the map from $\Ker\ D^\dag$ to $H^1(K)$ constructed above.

Since $\bD_p$ and multiplication by $z$ commute, the action of $\hphi$ on $H^1(K)$ is simply 
multiplication by $z$, without a need for a projection. This is the reason the cohomological
description of $\Ker\ D^\dag$ is useful.

\subsection{Explicit Argument}

Suppose the point $(\zeta, e^{2\pi\sigma})\in\CC\times\CC^*$ belongs to the Hitchin
spectral curve $\C$. In this case there exists a nonzero vector $\Theta\in H^1(K_\sigma)$ such that
\begin{equation} \label{hat}
\hphi(\sigma)\Theta=\zeta \Theta.
\end{equation}
As explained above, $\hphi$ acts on $H^1(K_\sigma)$ as multiplication by $z$. Let $\theta$ be a one-form 
representing $\Theta\in H^1(K_\sigma)$. Then the equation (\ref{hat}) means that there exists
$\psi\in \Lambda^{0,0}(X,E)$ such that 
\begin{equation}\label{exact}
(z-\zeta)\theta=\bD_1 \psi.
\end{equation}
It follows that $\bD_1 \psi$ vanishes at $z=\zeta$. In particular we have
\begin{equation}
\left(\frac{\partial}{\partial\chi}-iA_{\chi}-\phi+\sigma\right)\psi\vert_{z=\zeta}=0,
\end{equation}
i.e. the restriction of $\psi$ to the circle $z=\zeta$ is covariantly constant with respect
to the connection $B+i\sigma d\chi$. If $\psi$ is not identically zero on the circle
$z=\zeta$, this implies that
the holonomy matrix $V(\zeta,2\pi)$ has an eigenvalue equal to 
$e^{2\pi\sigma}$, and
consequently the point $(\zeta,e^{2\pi\sigma})$ belongs to the monopole spectral curve $\Sc$. 

To complete the proof of $\C=\Sc$ it remains to show that $\psi$ does not vanish identically
on the circle $z=\zeta$. Suppose it does vanish. Then (\ref{exact}) implies that on the circle
$z=\zeta$ we have $\frac{\partial^j}{\partial \bar{z}^j}\psi=0$ for all $j\geq 0$.
It follows that the function $\psi$ has the form $\psi=(z-\zeta) a(z,\bar{z}, \chi)+b(z,\bar{z}, \chi)$, 
where both $a$ and $b$ are smooth, and as $z\ra\zeta$ the function $b$ approaches zero faster 
than any power of $|z-\zeta|$. Hence $\psi$ is divisible by $(z-\zeta)$, i.e.
there exists a smooth function $\varphi\in \Lambda^{0,0}(X,E)$ such that $\psi=(z-\zeta)\varphi.$
Then $\theta=\bD_0\varphi$, which contradicts the assumption that $\theta$ represents
a nontrivial class in $H^1(K)$.

\subsection{Cohomological Argument}

Consider a complex of sheaves of vector spaces:
\begin{equation}\label{notexact}
0\ra E\xrightarrow{z-\zeta} E\xrightarrow{\ rest.\ }
E\vert_{z=\zeta}\xrightarrow{} 0,
\end{equation}
where the second map is multiplication by $z-\zeta$, and the map $rest.$ is restriction 
to the circle 
$z=\zeta$. This complex fails to be exact in the second term. Nevertheless, as shown below, 
there is a long exact sequence in $\bD$ cohomology:
\begin{equation}\label{cohomology}
0\ra H^0_\bD(\SS^1, E\vert_{z=\zeta})\ra H^1_\bD(X, E)
\xrightarrow{z-\zeta} H^1_\bD(X, E)
\xrightarrow{rest.} H_{\bD}^1(\SS^1, E\vert_{z=\zeta})\ra 0.
\end{equation}
Here $H^1_{\bD}(X,E)$ is the same as $H^1(K)$, while 
$H^j_\bD(\SS^1, E\vert_{z=\zeta})$ is the $j$-th cohomology of the 
restriction of $K$ to the circle $z=\zeta$.
Note that the restriction of $\bD_p$ to $z=\zeta$ is simply the covariant differential 
with respect to the connection $B+i\, s\, d\chi$ restricted to $z=\zeta$.

To understand where this exact sequence comes from, it is helpful to think about 
solutions of self-duality equations on $\CC\times T^2$. Periodic monopoles are a particular
class of such solutions which are invariant with respect to translations in one direction on 
the torus. Now the variable $\chi$ gets promoted to a complex variable parameterizing the 
universal cover of the torus, and the operator $\bD$ becomes simply a 
$\bpartial$--operator on the bundle $E$. Thus $E$ has a structure of a holomorphic bundle 
over $\CC\times T^2$. The cohomology of $\bD$ is the Dolbeault cohomology of $E$.
If we forget about noncompactness, the Dolbeault cohomology can be identified with the \v{C}ech
cohomology of the holomorphic bundle $E$. On the other hand, if we work in the holomorphic 
category, the sequence of sheaves (\ref{notexact}) is exact and hence
induces a long exact sequence of \v{C}ech cohomology groups.

In our situation, we cannot use the holomorphic interpretation. Instead, in the next subsection 
we derive the exact sequence~(\ref{cohomology}) from a spectral sequence of a double complex. 

The coincidence of the Hitchin and the monopole spectral data is an 
immediate consequence of
the exactness of the sequence~(\ref{cohomology}). Indeed, if the point 
$(\zeta,\exp(2\pi\sigma))$ 
belongs to
the Hitchin spectral curve $\C$, then the kernel of the map $(z-\zeta)$ from 
$H^1(K_\sigma)$ to $H^1(K_\sigma)$ is nontrivial. But the cohomology exact sequence implies 
an isomorphism
\begin{equation}\label{kernel}
\Ker(z-\zeta)\cong H_{\bD}^0(\SS^1, E\vert_{z=\zeta}).
\end{equation}
Therefore  $H_{\bD}^0(\SS^1, E\vert_{z=\zeta})$ is nontrivial as well. This means that the
holonomy of $B$ along the circle $z=\zeta$ has $\exp(2\pi\sigma)$ as one of its eigenvalues. 
Thus the point $(\zeta, \exp(2\pi\sigma))\in \CC\times \CC^*$ belongs to the monopole
spectral curve $\Sc$.
Moreover, the fibers of the spectral line bundles on $\C$ and $\Sc$ are given
by $\Ker(z-\zeta)$ and $H_{\bD}^0(\SS^1, E\vert_{z=\zeta}),$ respectively. Thus we also get
an isomorphism of the line bundles. 

\subsection{Exactness Of The Cohomology Sequence}\label{sec:Exactness}

Consider again the complex of sheaves
\begin{equation}\label{three}
0\ra E\xrightarrow{(z-\zeta)} E\xrightarrow{rest.}
E\vert_{z=\zeta}\ra 0,
\end{equation}
where $rest.$ is the restriction to $z=\zeta$. Since $\bD_p,\ p=0,1,$ commutes with
$(z-\zeta)$ and $rest.$, this complex is included in a double complex $D^{p,q}$:
\begin{equation}
\nonumber
\xymatrix{
&&E\otimes\Lambda^{0,2} \ar[r]^{(z-\zeta)} & E\otimes\Lambda^{0,2} 
\ar[r]^-{rest.} & 0\\
D^{p,q}: &&E\otimes\Lambda^{0,1}
\ar[r]^{(z-\zeta)}\ar[u]^{\bD_1}&E\otimes\Lambda^{0,1} 
\ar[r]^-{rest.}\ar[u]^{\bD_1}&E\otimes\Lambda^{0,1}\vert_{z=\zeta}
\ar[u]^{\bD_1}\\
&&E\otimes\Lambda^{0,0} \ar[r]^{(z-\zeta)}\ar[u]^{\bD_0}& 
E\otimes\Lambda^{0,0} 
\ar[r]^-{rest.}\ar[u]^{\bD_0} &E\otimes\Lambda^{0,0}\vert_{z=\zeta}
\ar[u]^{\bD_0}.
}
\end{equation}

Computing the cohomology of the rows, we obtain the first term 
of the ``vertical'' spectral sequence :
\begin{equation}
\nonumber
\xymatrix{
&&0 &  {\left\{\eta^{0,2}\sim\eta^{0,2}+(z-\zeta)\omega^{0,2}\right\}}& 0\\
E_1^{p,q}:&&0 \ar[u] & {\left\{\eta^{0,1}\sim\eta^{0,1}+(z-\zeta)\omega^{0,1}\vert 
\ rest.(\eta^{0,1})=0\right\}}\ar[u]^{\bD_1}&0\ar[u]\\
&&0 \ar[u]&{\left\{\eta^{0,0}\sim\eta^{0,0}+(z-\zeta)\omega^{0,0}\vert 
\ \eta^{0,0}|_{z=\zeta}=0\right\}}\ar[u]^{\bD_0}&0\ar[u].
}
\end{equation}
On the second level the ``vertical'' spectral sequence degenerates to zero: 
$E_{\infty}^{p,q}=0$.

Now let us compute the ``horizontal'' spectral sequence. Its first term is simply 
the $\bD$ cohomology of $D^{p,q}$: $\tE_1^{p,q}=H_{\bD}(D^{p,q})$. The second term 
$\tE_2^{p,q}$ is given by
\begin{equation}
\nonumber
\xymatrix{
0 \ar[drr]^{d_2} & 0  & 0 \\
\Ker\ (z-\zeta)\vert_{H^1(K)}\ar[drr]^{d_2} & \Ker\ rest./{\rm Im}\ (z-\zeta)\vert_{H^1(K)} 
& H_{\bD}^1(E\vert_{z=\zeta})/{\rm Im}\ rest. \vert_{H^1(K)}\\
0 &  0 & H_{\bD}^0\left(E\vert_{z=\zeta}\right) .
}
\end{equation}
At the next level the ``horizontal'' spectral sequence degenerates. Since the 
``vertical'' spectral 
sequence converges to zero, so should the ``horizontal'' one. 
%Comparing $\tE_3^{p,q}$ 
%\begin{equation}
%\nonumber
%\xymatrix{
%&0 & 0  & {\rm Im}_{d_2}\\
%\tE_3^{p,q}&\Ker_{d_2} & \Ker_{restr.}H^1(E)/(z-\zeta) H^1(E) 
%& H^1(E\vert_{z=\zeta})/{\rm Im}_{\delta} H^1(E)\\
%&0 &  0 &  0.
%}
%\end{equation}
{}From this we infer the isomorphisms 
\begin{align}
\Ker\ rest.\vert_{H^1(K)} & \cong{\rm Im}\ (z-\zeta)\vert_{H^1(K)},\nn\\  
H_{\bD}^1(E\vert_{z=\zeta}) & \cong{\rm Im}\ rest.\vert_{H^1(K)},\nn\\ 
\Ker\ (z-\zeta)\vert_{H^1(K)} & \cong H_{\bD}^0\left(E\vert_{z=\zeta}\right).\nn
\end{align}
These isomorphisms are equivalent to the exactness of the cohomology 
sequence~(\ref{cohomology}).

\subsection{Revisiting The Index Computation}

Using the spectral sequence technology, we can give another proof that $\dim H^1(K)=k.$ The advantage of this method of proof is that it makes
it clear that the zero modes of the Dirac operator are ``localized''
near the points $z=\zeta_1,\ldots,\zeta_k$.

Consider a bundle morphism $\Delta$ from $E$ to $E$ defined as multiplication by the 
polynomial
$$\det(z-\hphi|_{s=0})=(z-\zeta_1)\ldots (z-\zeta_k).$$ Here $\zeta_1,\ldots,\zeta_k$
are the roots of the characteristic polynomial of $\hphi$, or equivalently the solutions
of the equation $\Tr\ V(z,2\pi)=2.$ Suppose all $\zeta_i$ are distinct. Let $Y$ be the restriction of the sheaf 
$E$ to the union of $k$ circles $z=\zeta_1,\ldots,z=\zeta_k$, and let 
$rest.$ be the restriction map.
Obviously, $rest.\cdot\Delta=0$, so we get a complex
$$
0\ra E\xrightarrow{\Delta} E\xrightarrow{rest.} Y\ra 0.
$$
As before, this complex is not exact in the middle term, but nevertheless leads to an exact 
cohomology sequence:
$$
0\ra H_{\bD}^0(Y)\xrightarrow{}H^1(K)\xrightarrow{\Delta} H^1(K)\xrightarrow{rest.}H_{\bD}^1(Y)\ra 0.
$$
The proof of exactness is identical to the one given above. 
Now note that the map $\Delta$ sends $H^1(K)$ to zero by virtue of the Cayley-Hamilton theorem on
the characteristic polynomial of a matrix. Hence $H^1(K)$ is isomorphic to $H_{\bD}^0(Y)$.
On the other hand, if all the numbers $\zeta_\alpha$ are distinct, we have
$$
H_{\bD}^0(Y)=\oplus_{\alpha=1}^k H_{\bD}^0\left(E\vert_{z=\zeta_\alpha}\right)\cong \CC^k.
$$
Hence $\dim H^1(K)=k$. Moreover, we see that each of the circles $z=\zeta_\alpha$
gives rise to a vector in $H^1(K)\cong\Ker\ D^\dag$, and all these vectors are linearly
independent. Thus we may think of the $k$ zero modes of the Dirac operator $D^\dag$ as ``localized''
in the neighborhood of $k$ circles $z=\zeta_\alpha,\ \alpha=1,\ldots,k.$

\section{Asymptotic Behavior Of The Hitchin Data}\label{sec:asymptotics}

The fact that the spectral curves $\C$ and $\Sc$ coincide provides a wealth of information
about periodic monopoles. In particular, it allows to determine the
behavior of the Higgs field $\hphi$ for ${\rm Re}\ s\ra\pm\infty.$

Let us rewrite the equation of the curve $\Sc$ in the following form:
\begin{equation}\label{curvetwo}
f(z)-\exp(2\pi s)-\exp(-2\pi s)=0.
\end{equation}
Here we used the identification of the eigenvalue $w$ of $V(z,2\pi)$ with $\exp(2\pi s)$.
Recalling the definition of the Hitchin spectral curve $\C$, we infer that
all coefficients of the characteristic polynomial of $\hphi$ except $\det\hphi$ are 
independent of $s$. Furthermore, we know from section~\ref{sec:specS} that $f(z)$ 
is a degree $k$ polynomial in $z$ with leading coefficient $\exp(2\pi\ups)$.
It follows that the determinant of $\hphi$ is given by
$$
\det\hphi=(-1)^{k+1}\left(e^{2\pi s}+e^{-2\pi s}\right)\exp(-2\pi\ups).
$$

Another piece of information comes from the equation~(\ref{FHitchin}) which says that
the curvature of $\hA$ is proportional to the restriction
of the operator $(D^\dag D)^{-1}$ to the subspace $\Ker\ D^\dag$. The operator 
$(D^\dag D)^{-1}$ is a integral operator on the space of $L^2$ sections of 
$E\ot S$, and from (\ref{Wone}) it is clear that its norm vanishes in the limit 
${\rm Re}\ s\ra\pm\infty$. (In fact, it is easy to see that the norm is bounded from above
by a multiple of $1/|{\rm Re}\, s|^{3/2}.$) Hence the curvature of $\hA$ also goes to 
zero in this limit. 
Bogomolny equations then imply that $[\hphi^\dag,\hphi]\ra 0$
asymptotically. 

Combining this with the information about the characteristic polynomial of $\hphi$, we 
infer that for ${\rm Re}\ s\ra\pm\infty$ the Higgs field $\hphi$ behaves as follows:
\begin{equation}\label{asympttwo}
\hphi(s)\sim -\exp\left(-\frac{2\pi}{k}\left(\ups+\frac{i}{2}\right)\right)
\cdot g_{\pm}(s)\ e^{\pm \frac{2\pi s}{k}}\ (1+o(1)) 
\diag(1,\omega,\omega^2,\ldots,\omega^{k-1})\ g_{\pm}(s)^{-1}.
\end{equation}
Here $\omega$ is a $k$--th root of unity and $g_\pm(s)$ are multi-valued functions on $\hX$ with 
values in $U(k)$. In order for $\hphi$ to be well-defined, the functions $g_\pm$ must
satisfy
$$
g_\pm(s+i)=g_\pm(s)\ V^{\pm 1}e^{i\beta}.
$$
Here $\beta$ is a real number, and $V\in SU(k)$ is the so-called
``shift'' matrix:
\begin{equation}\label{shift}
\qquad V=
\begin{pmatrix}
0 & 0 & 0 & \dots & 0 & 1 \\
1 & 0 & 0 & \dots & 0 & 0 \\
0 & 1 & 0 & \dots & 0 & 0 \\
\hdotsfor{6} \\
0 & 0 & 0 & \dots & 0 & 0 \\
0 & 0 & 0 & \dots & 1 & 0 \\
\end{pmatrix}.
\end{equation}

We can reformulate these results as follows. The Nahm transform of a periodic monopole 
of charge $k$ is a pair 
$(\hA,\hphi)$ satisfying the $U(k)$ Hitchin equations and the following asymptotic
conditions:

(i) The functions $\Tr\ \hphi(s)^\alpha,\ \alpha=1,\ldots,k-1$ are bounded;

(ii) The function $\exp(\mp 2\pi s)\det\ \hphi(s)$ behaves as 
$$(-1)^{k+1}\exp(-2\pi\ups)+O(\exp(\mp 2\pi s))$$ for 
${\rm Re}\ s\ra\pm\infty$;

(iii) $\displaystyle ||F_{z\bz}||^2\leq \frac{C}{|{\rm Re}\, s|^3}.$

Since the functions $\Tr\ \hphi(s)^\alpha$ and $\det \hphi(s)$ are holomorphic functions
on $\CC^*\cong\RR\times\SS^1$ by virtue of the Hitchin equations, the first two conditions 
are equivalent to the statement that the spectral curve of $(\hA,\hphi)$ has the 
form~(\ref{curvetwo}), with $f(z)$ being a polynomial of degree $k$ with 
the leading coefficient $\exp(2\pi\ups)$, and the rest of the coefficients being arbitrary constants.

In the next three sections we will show that the correspondence between the
solutions of Hitchin equations satisfying (i)-(iii) and periodic monopoles is one-to-one.

\section{The Inverse Nahm Transform}\label{sec:Inverse}

In this section we show how to associate a periodic $SU(2)$ monopole of charge $k$
to any solution of $U(k)$ Hitchin equations on $\hX$ with asymptotic
behavior as above. This procedure will be called the inverse Nahm transform.
It will take us from Hitchin data associated with a bundle $\hE\rightarrow\hX$
to monopole data on a bundle $\check{\hE}\rightarrow X$. Later on, in section~\ref{sec:Nahmsq},
we will show that the monopole on $\check{\hE}$ coincides with that on $E$, so in this section
we shall use a simplified notation in which the symbol~$\check{\hat{}}$~is omitted.
Since the original monopole data on $E$ are not used in this section, this should not lead to
confusion.

Let $\hE$ be a trivial unitary rank $k$ bundle over $\hX=\CC^*$,
$\hA$ be a connection on $\hE$ and $\hphi$ be a section of  $\End\ (\hE)$.
{}Furthermore, let the pair $(\hA,\hphi)$ be a solution of $U(k)$ Hitchin equations on $\hX$ such that $\hphi$ has the asymptotics
as in ~(\ref{hasympt}). Let $\hL$ be a trivial line bundle over $\hX$ with a flat unitary connection $\ha$
such that the holonomy of $\ha$ around the positively oriented loop encircling the 
origin of $\CC^*$ is $\exp(-i\chi).$ The variable $\chi$ is assumed
to take values in the interval $[0,2\pi].$
Consider a Dirac--type operator $\hD: \hE\ot\hL\ot\CC^2\to\hE\ot\hL\ot\CC^2$ given
by
$$
\hD=\begin{pmatrix}
-\hphi+z & 2\partial_{\hA+\ha} \\
2\bpartial_{\hA+\ha} & -\hphi^\dag+\bz .
\end{pmatrix}
$$
Here $z$ is a complex parameter.
The operator $\hD$ is Fredholm for any $z$ and $\chi$ because $||\hphi||$ grows
without bound as $t\ra\pm\infty$.

The Weitzenbock formula for $\hD$ reads:
$$
\hD^\dag\hD=\begin{pmatrix}
(\hphi^\dag-\bz)(\hphi-z)-4\partial_{\hA+\ha}\bpartial_{\hA+\ha} & 2\partial_{\hA}\hphi^\dag \\
2\bpartial_{\hA}\hphi & (\hphi-z)(\hphi^\dag-\bz)-4\bpartial_{\hA+\ha}\partial_{\hA+\ha}.
\end{pmatrix}
$$
If the Hitchin equations are satisfied, then this formula simplifies:
$$
\hD^\dag\hD=-\nabla_{\hA+\ha}^2+\frac12((\hphi^\dag-\bz)(\hphi-z)+
(\hphi-z)(\hphi^\dag-\bz)).
$$ 
This operator is clearly positive definite on the space of smooth rapidly decreasing 
sections of $\hE\ot\hL\ot\CC^2$. It is easy to see that any $L^2$ eigenvector of $\hD$ with
zero eigenvalue must be smooth and decreasing faster than any negative power of $r={\rm Re}\ s$,
hence $\hD$ has trivial $L^2$ kernel. Thus the dimension of the kernel of $\hD^\dag$
is minus the index of $\hD$.

Computing the $L^2$ index of $\hD$ turns out to
be rather tricky. We will do it in the next section by reinterpreting $\Ker\ \hD^\dag$ as a 
certain cohomology group and computing it using the spectral sequence of a double complex, 
similarly to how it was done in Section~\ref{sec:CandS}. The result of this computation is that 
$\Ker\ \hD^\dag$ has dimension $2$ for all $z$ and $\chi$.

We conclude that $\Ker\ \hD^\dag$ forms a trivial rank $2$ bundle on the manifold 
$X=\CC\times\SS^1$ parameterized
by $(z,\chi)$. Since the elements of the kernel are square-integrable,
we have a well-defined Hermitian inner product
on $\Ker\ \hD^\dag$ for all $z,\chi$. In this way we obtain a unitary rank 2 bundle
$E$ over $X$.

Now we need to define a connection $A$ on $E$ and a traceless Hermitian
section $\phi$ of $\End (E)$.
The connection on $\Ker \hD$ is induced from a trivial connection on a trivial 
infinite-dimensional bundle on $X$ whose fiber consists of all smooth $L^2$ sections of $\hE\ot \hS\ot \hL$. If we introduce the projectors $\hP=1-\hD (\hD^\dag \hD)^{-1} 
\hD^\dag$ and $\hQ=1-\hP$, we may write 
$$
d_A=\hP d=\hP\left(dz \frac{\partial}{\partial z}+
d\bz\frac{\partial}{\partial \bz}+d\chi \frac{\partial}{\partial\chi}\right).
$$
The value of the Higgs field $\phi$ at a point $x\in X$ is a linear map
$E\ra E$ defined as a composition of multiplication by $r$ and projection to
$\Ker D^\dag$, i.e. 
$$
\phi=\hP r.
$$

It remains to show that $\phi$ and $A$ satisfy the Bogomolny
equation~(\ref{Bogomolny}).
In the coordinates $z,\chi$ used above this equation is equivalent
to a pair of equations
\begin{eqnarray}
&&F_{\bz\chi}=i\bpartial_A\phi, \\
&&F_{z\bz}=\frac{i}{2}\left(i_{\partial/\partial\chi}\cdot d_A\right)\phi.
\end{eqnarray}
Here $\bpartial_A$ means $i_{\partial/\partial\bz}\cdot d_A.$

To show that these equations are satisfied, we have to use the commutation relations
$$
\begin{array}{llll}
{}\dst{[\hD,\frac{\partial}{\partial\chi}]=-i\sigma_2}, &
\dst{[\hD,\frac{\partial}{\partial z}]=-p_+}, & 
\dst{[\hD,\frac{\partial}{\partial\bz}]=-p_-}, & 
\dst{[\hD,r]=\sigma_1}, \\
{}\dst{[\hD^\dag,\frac{\partial}{\partial\chi}]=i\sigma_2}, &
\dst{[\hD^\dag,\frac{\partial}{\partial z}]=-p_-}, & 
\dst{[\hD^\dag,\frac{\partial}{\partial\bz}]=
-p_+}, & \dst{[\hD^\dag,r]=-\sigma_1},
\end{array}
$$
and the fact that $\hD^\dag \hD$ commutes with all $\sigma_i$. 
We find for the curvature of $A$:
\begin{equation}\label{curvatures}
\begin{array}{rcl}
F_{\bz\chi} & = & i\hP (\bpartial \hQ\pchi\hQ-\pchi\hQ\bpartial \hQ)\\
 & = & i\hP([\bpartial,\hD](\hD^\dag\hD)^{-1}[\pchi,\hD^\dag]-
 [\pchi,\hD](\hD^\dag\hD)^{-1}[\bpartial,\hD^\dag])\\
 & = & \hP(p_-(\hD^\dag\hD)^{-1}\sigma_2+\sigma_2(\hD^\dag\hD)^{-1}p_+)\\
 & = & 2i\hP(\hD^\dag\hD)^{-1}\sigma_-,\\
F_{z\bz} & = & i \hP(\partial\hQ\bpartial\hQ-\bpartial\hQ\partial\hQ)\\
 & = & i \hP([\partial,\hD](\hD^\dag\hD)^{-1}[\bpartial,\hD^\dag]-
[\bpartial,\hD](\hD^\dag\hD)^{-1}[\partial,\hD^\dag])\\
 & = & i\hP(p_+(\hD^\dag\hD)^{-1}p_+ - p_-(\hD^\dag\hD)^{-1}p_-)\\
 & = & i\hP(\hD^\dag\hD)^{-1}\sigma_3.
\end{array}
\end{equation}
The covariant derivatives of $\phi$ can also be easily computed:
\begin{equation}\label{derivatives}
\begin{array}{rcl}
\bpartial_A\phi & = & [\hP\bpartial,\hP r]\\
 & = & \hP r\hQ\bpartial-\hP\bpartial\hQ r\\
 & = & \hP([r,\hD](\hD^\dag\hD)^{-1}[\hD^\dag,\bpartial]-[\bpartial,\hD]
 (\hD^\dag\hD)^{-1}[\hD^\dag,r])\\
 & = & \hP (\sigma_1(\hD^\dag\hD)^{-1}p_+ + p_-(\hD^\dag\hD)^{-1}\sigma_1)\\
 & = & 2\hP(\hD^\dag\hD)^{-1}\sigma_-,\\
(i_{\partial/\partial\chi}\cdot d_A)\phi & = & [\hP\pchi,\hP r]\\
& = & \hP r\hQ\pchi-\hP\pchi\hQ r\\
& = & \hP([r,\hD](\hD^\dag\hD)^{-1}[\hD^\dag,\pchi]-[\pchi,\hD](\hD^\dag\hD)^{-1}
[\hD^\dag,r])\\
& = & -i\hP(\sigma_1(\hD^\dag\hD)^{-1}\sigma_2-
\sigma_2(\hD^\dag\hD)^{-1}\sigma_1)\\
& = & 2 \hP(\hD^\dag\hD)^{-1}\sigma_3.
\end{array}
\end{equation}
Comparing (\ref{curvatures}) and (\ref{derivatives}), we see that $A$ and $\phi$
indeed satisfy the Bogomolny equation.

\section{Spectral Data And The Inverse Nahm Transform}\label{sec:invspec}

In the previous section we showed that the inverse Nahm transform applied to 
a solution of Hitchin equations on $\hX\cong \CC^*$ yields a solution 
of the Bogomolny equation on $X\cong \RR^2\times \SS^1$. In this section
we prove that if the solution of the Hitchin equations has the 
asymptotics~(\ref{asympttwo}), then the corresponding solution of the Bogomolny
equation has the asymptotics~(\ref{asympt}). The use of the monopole spectral
data greatly facilitates this proof. We first give a ``pedestrian''
proof of the coincidence of the Hitchin and monopole spectral data,
and then a more conceptual one using the spectral sequence of a double
complex. This spectral sequence will also be used to compute the index
of the Dirac operator $\hD$, thereby filling a gap in the derivation 
of Section~\ref{sec:Inverse}. In fact, we will show that the zero modes
of the twisted Dirac operator $\hD_{z,\chi}$ are in one-to-one
correspondence with the points on $\hX$ where $\hphi$ has an eigenvalue
$z$.

\subsection{Cohomological Interpretation Of The Inverse Nahm Transform}

In order to give a cohomological interpretation of the inverse Nahm transform,
let us consider the following complex constructed from the sheaves of vector
spaces $\hLambda^0=\Lambda^{0,0}(\hX,\hE)$ and $\hLambda^1=\Lambda^{0,1}(\hX,\hE)$: 
\begin{equation}\label{thehat}
0\ra \hLambda^0 \xrightarrow{\hdelta_0} \hLambda^0\oplus\hLambda^1 
\xrightarrow{\hdelta_1}
\hLambda^1 \ra 0,
\end{equation}
with $\hdelta_0$ and $\hdelta_0$ defined by
\begin{equation}\label{deltahat}
{\hdelta_0}: f \mapsto \begin{pmatrix} 
                          -(\hphi-\zeta) f\\ 2\bpartial_{\hA+\ha} f
                   \end{pmatrix},\ \ 
\hdelta_1:\begin{pmatrix} g_0\\ g_1 \end{pmatrix}
\mapsto \left(-2\bpartial_{\hA+\ha} g_0-(\hphi-\zeta) g_1\right).
\end{equation}
Since every element $g_1\in\hLambda^1$ has the form $g_1=g(s) d\bar{s}$, 
we can identify $\hLambda^0\ni g$ with $\hLambda^1\ni g_1$.

In terms of this complex $\hK_{z,\chi}$ the Dirac operator $\hDd$ of 
section~\ref{sec:Inverse} is given by
\begin{equation}
\hDd=\hdelta^*_0-\hdelta_1.
\end{equation}
One can easily see that the square-integrable zero modes of $\hDd$ are in one-to-one
correspondence with the first $L^2$ cohomology of the above complex. 
Thus, if we denote the inverse Nahm transform of the
bundle $\hE$ by $\chE$, we have a canonical identification of
the fiber of $\chE$ at a point $(z,\chi)\in X$ with 
$H^1(\hK_{z,\chi})$. In other words, the spaces $H^1(\hK_{z,\chi})$
form a vector bundle over $X$ which is canonically
isomorphic to $\chE$. 

At this stage of the discussion it becomes crucial to keep track of the
periodicity conditions along the $\chi$ and $t$ directions.
Let us denote the circles parameterized by $\chi$ and $t$ by $\SS^1$ and
$\hSS^1$, respectively.
Previously we worked with one circle at a time
and could choose a trivialization of any bundle on a circle
so that the components of a section be periodic functions.
However, if one considers a bundle on $\SS^1\times\hSS^1$, 
both circles are in the game. 
If the bundle on $\SS^1\times\hSS^1$ is nontrivial,
then there is no trivialization in which sections are periodic functions
along both periodic directions. This is in fact what happens in our case.

Let us introduce some notation. The circle $\hSS^1$ parameterizes
line bundles with a unitary connection on $\SS^1$. Namely,
a point $t\in\hSS^1$ corresponds to a line bundle with a monodromy
$e^{-2\pi i t}$. We will denote this line bundle with a connection
by $L_t$. If one chooses a ``periodic'' trivialization alluded to
above, then the connection on $L_t$ is $-td\chi$. Alternatively,
if one chooses a ``quasiperiodic'' trivialization in which sections are 
represented by functions satisfying $f(\chi+2\pi)=e^{2\pi i t} f(\chi)$,
then the connection on $L_t$ is trivial. Conversely, $\SS^1$ parameterizes
unitary line bundles on $\hSS^1$; the point $\chi\in \SS^1$ corresponds
to a line bundle $\hL_\chi$ whose monodromy is $e^{i\chi}$. 

Recall now the definition of the Poincar\'e line bundle $\cP$ on 
$\SS^1\times \hSS^1$~(see e.g.~\cite{DK}). It is a line bundle 
with a unitary connection whose restriction to any circle $t=t_0$ is 
isomorphic to $L_{t_0}$ (as a line bundle with a connection), while
its restriction to any circle $\chi=\chi_0$ is isomorphic to 
$\hL_{\chi_0}$. The curvature of the connection on $\cP$ is
given by $d\chi\wedge dt$. The Poincar\'e line bundle 
is nontrivial and has the first Chern class equal to $1$.
If we choose a trivialization of $\cP$
such that the connection on $\cP$ is given by
$\cA=\chi dt$, then its sections are represented by functions
$f(\chi,t)$ satisfying $f(\chi+2\pi,t)=e^{2\pi i t} f(\chi,t)$
and $f(\chi,t+1)=f(\chi,t)$. Similarly, the dual line bundle $\cP^*$
has a connection 1-form $-\chi dt$, and its sections are represented by
functions $f(\chi,t)$ satisfying $f(\chi+2\pi,t)=e^{-2\pi i t} f(\chi,t)$
and $f(\chi,t+1)=f(\chi,t)$.

By making use
of $\cP^*$, the inverse Nahm transform  can be rephrased as follows. We pull back
the bundle $\hE$ to $X\times\hX$ using the natural projection
$\hpi: X\times \hX\ra \hX$. Then we twist $\hpi^*(\hE)$ by a line bundle with
a unitary connection $-\chi dt$ and trivial periodicity condition in the
$t$ direction, i.e. by a line bundle $\cP^*$.
The complex $\hK$ is also twisted by $\cP^*$, as is clear from its definition,
and in addition by the Higgs field $z\, ds$.
{}Finally, we form a bundle on $X$ whose fiber over $x$ is the
first cohomology group of the complex $\hK_{z,\chi}$. The direct Nahm transform
can be similarly reformulated using the line bundle $\cP$.
(This description suggests that it is useful to think about the derived
functor of the Nahm transform, which reduces to the ordinary Nahm transform 
when both the initial and the transformed complexes happen to have only
a single nonvanishing cohomology).

% An algebraic geometer
%would say that $\chE$ is given by $R^1\pi_*(\cP^*\otimes\hpi^*(\hE))$, where
%$R^1\pi_*$ is the higher direct image functor (with respect to the
%differential $(\hdelta_0,\hdelta_1)$).

The upshot of this discussion is that sections of $\cP^*\times\hE$
should be thought of as vector-valued functions on $X\times\hX$ satisfying 
$$e^{2\pi it} f(\chi+2\pi,t)=f(\chi,t),\quad f(\chi,t+1)=f(\chi,t),$$
while the covariant derivatives along $\frac{\partial}{\partial\chi}$
and $\frac{\partial}{\partial t}$ are given by
$$
\nabla_\chi=\frac{\partial}{\partial\chi}, \quad
\nabla_t=\frac{\partial}{\partial t}+i\chi-i\hA_t.
$$ 

\subsection{Coincidence Of The Spectral Curves:  An Explicit Argument}
Let us pick a point $(\zeta, \exp{2\pi \sigma})\in\CC\times\CC^*$ belonging to the 
monopole spectral curve $\Sc$. In other words, $\exp{2\pi\sigma}$ is one of the 
eigenvalues of the holonomy $V(\zeta,2\pi)$:
\begin{equation}
\det(V(\zeta,2\pi)-e^{2\pi\sigma})=0.
\end{equation}
This implies that there is a family of sections $\Psi$ of $\hLambda^0\oplus\hLambda^1$  
parameterized by $\chi\in \RR$,
such that $[\Psi(\chi)]$ is a nonzero element of $H^1(\hK_{z,\chi})$ and 
\begin{equation}\label{Psieq}
\left[\left(\frac{\partial}{\partial\chi}-r+\sigma\right) \Psi\right]=[0].
\end{equation}
Here the brackets designate the cohomology class in $H^1(\hK_{z,\chi})$.
We also denote $ s=r+i t$, as usual.
 
Let us unwrap the equation~(\ref{Psieq}). We will write $\Psi$ as follows:
$$
\Psi(s,\chi)=\begin{pmatrix} a(\chi,s)\\b(\chi,s)\end{pmatrix},\quad
$$
{}For a fixed $\chi$ the functions $a$ and $b$ are sections of $\hE$. We can choose the cohomology 
representative $\Psi(s,\chi)$ so that $a$ and $b$ 
are sections of $\cP^*\ot\hpi^*(\hE)$, and therefore satisfy
$$
e^{2\pi it} \begin{pmatrix} a(\chi+2\pi,s) \\b(\chi+2\pi,s)\end{pmatrix}=
\begin{pmatrix} a(\chi,s)\\b(\chi,s)\end{pmatrix}.
$$
The equation~(\ref{Psieq}) means that there exists a section $h$ of 
$\cP^*\ot\hpi^*(\hE)$
such that
$$
\left(\frac{\partial}{\partial\chi}-r+\sigma\right) 
\begin{pmatrix} a\\b\end{pmatrix}=
\begin{pmatrix} -(\hphi-\zeta)h\\2\bpartial_{\hA+\ha} h\end{pmatrix}.
$$
Introducing
$$
F(s)=\int_0^{2\pi} e^{i\chi t} h(\chi,s) d\chi,
$$
we find that
$$
\left(\hphi(\sigma)-\zeta\right) F(\sigma)=0.
$$
If $F(\sigma)$ is nonzero, this implies that $\zeta$ 
is an eigenvalue of $\hphi(\sigma)$ and therefore
the point $(\zeta,e^{2\pi\sigma})$ belongs to the Hitchin spectral curve $\C$.
This shows that the curves coincide.

To prove that $F(\sigma)$ is nonzero, let us assume the contrary. Then it follows
from the Fredholm Alternative that the equation 
$$
\left(\frac{\partial}{\partial\chi}-r+\sigma\right)g=h
$$
has a solution $g\in\Gamma(\cP^*\ot\hpi^*(\hE))$.  One can easily see that 
$a$ and $b$ are expressible in terms of $g$ as follows:
$$
\begin{pmatrix} a\\b\end{pmatrix}=
\begin{pmatrix} -(\hphi-\zeta)g\\2\bpartial_{\hA+\ha} g\end{pmatrix}.
$$
But this contradicts the assumption that $\Psi$ represents a nontrivial
cohomology class in $H^1(\hK_{z,\chi}).$ Thus if $\exp(2\pi\sigma)$ is an
eigenvalue of $V(\zeta,2\pi)$, then $\zeta$ is an eigenvalue of $\hphi(\sigma)$.

\subsection{Cohomological Argument}

The above argument can be conveniently rephrased in cohomological terms.
Consider the manifold 
$\SS^1\times\hX=\SS^1\times\hSS^1\times\RR$ and a bundle $\E=\cP^*\ot\hpi^*(\hE)$
over it. We have two operators acting on its sections,
namely $\hdelta_{\zeta,\chi}$ of Eq.~(\ref{deltahat}) and $\daleth$ given by
\begin{equation}
\daleth=\frac{\partial}{\partial\chi}-r+\sigma.
\end{equation}
Let us note that these operators, $\hdelta_{\zeta,\chi}$ and $\daleth$, commute.
\begin{equation}
\left[\hdelta_{\zeta,\chi}, \daleth\right]=0.
\end{equation}

Consider a complex of sheaves of vector spaces:
\begin{equation}\label{leo}
0\rightarrow \E \xrightarrow{\daleth}\E\xrightarrow{\rm int} \hE_{s=\sigma}\rightarrow 0,
\end{equation}
where ${\rm int}$ acts as
\begin{equation}
{\rm int} : f(\chi,s) \mapsto\int_0^{2\pi}f(\chi,\sigma)e^{i\chi\, {\rm Im}\sigma} 
d\chi\end{equation}

Even though this short sequence is not exact, by an argument similar to that in 
subsection~\ref{sec:Exactness}, one can show that
there still is a long exact sequence of cohomology groups. To show this, we consider a double 
complex whose lowest row is the complex~(\ref{leo}), and the vertical differential
is given by $\hdelta_{\zeta,\chi}$.
The first level 
$\tE_1^{p,q}$ of the spectral sequence of this double complex
contains the cohomology groups $H_{\hdelta_{\zeta,\chi}}^j(\hX,\hE)$, as well as maps 
between them. Comparison with the total
cohomology of this double complex provides us with an exact sequence
\begin{equation}
0\rightarrow H^0_{\hdelta_{\zeta,\chi}} (\hE\vert_{s=\sigma})\rightarrow 
H^1_{\hdelta_{\zeta,\chi}}(\hX,\hE)\xrightarrow{\daleth}
H^1_{\hdelta_{\zeta,\chi}}(\hX,\hE)\rightarrow \ldots
\end{equation}
{}From the definition of $\hdelta_{\zeta,\chi}$ we have $H^0_{\hdelta_{\zeta,\chi}} 
(\hE\vert_{s=\sigma})=\Ker \left(\hphi(\sigma)-\zeta\right)$.
Recalling the identification of $\ \Ker\ \hDd_{\zeta,\chi}$ with 
$H^1_{\hdelta_{\zeta,\chi}}(\hX,\hE)$, the above exact 
sequence implies an isomorphism
\begin{equation}\label{another}
\Ker \left(\hphi(\sigma)-\zeta\right)\cong \Ker\ \daleth\vert_{\Ker \hDd_{\zeta,\chi}}.
\end{equation}
If the point $(\zeta,\exp(2\pi\sigma)$ belongs to the monopole spectral curve $\Sc$, then the 
right-hand side of this equation is nonempty, and therefore the point belongs to the
Hitchin spectral curve as well. The converse statement is also true. Thus the two curves coincide. 
Moreover, the spectral line bundle
$\Ker\left(\hphi(\sigma)-\zeta\right)$ on $\C$ is identified with
$\Ker\left(\daleth|_{H^1_{\hdelta_{\zeta,\chi}}(\hX,\hE)}\right)$, which is the line bundle on
the monopole spectral curve $\Sc$.
Thus the line bundles on $\C$ and $\Sc$ are also isomorphic.

The isomorphism~(\ref{another}) also enables one to compute the index of $\hD$.
Indeed, since for $\sigma_1\neq\sigma_2$ the kernels of $\daleth_{\sigma_1}$
and $\daleth_{\sigma_2}$ do not intersect, one can easily see that
$$
\Ker\ \hDd_{\zeta,\chi}\cong\oplus_{\sigma\in \hX} \Ker \left(\hphi(\sigma)-\zeta\right). 
$$
The dimension of the right-hand side is just
the number of points at which the spectral curve $\C$
intersects the cylinder in $\CC\times\CC^*$ given by $z=\zeta$. From the equation of
the curve (\ref{curvetwo}) we see that there are two such points for any $\zeta$. 
Since the kernel of $\hD$ is trivial, the index of $\hD$ equals $-2$, as promised.

\subsection{The Asymptotic Behavior Of The Monopole}

We are now ready to show that the inverse Nahm transform produces
a solution of Bogomolny equations with the asymptotics~(\ref{asympt}).
By assumption, we started from a solution of Hitchin equations
with the spectral curve
$$
f(z)-\exp(2\pi s)-\exp(-2\pi s)=0,
$$
where $f(z)$ is a degree $k$ polynomial with leading coefficient
$\exp(2\pi\ups)$.
We proved that the monopole spectral curve is given by the same
equation. This implies that $\Tr\ V(z,2\pi)$ grows as 
\begin{equation}\label{trgrowth}
\Tr\ V(z,2\pi)\sim z^k e^{2\pi\ups}
\end{equation}
for large $z$, while $\det V(z,2\pi)=1$ everywhere. 

Another piece of information that we need is that $||F_A||^2$ is
bounded by a multiple of $1/|z|^2$. This follows from~(\ref{curvatures})
and a simple estimate of the norm of $(\hD^\dag\hD)^{-1}$.
Together with the Bogomolny equation this fact implies that 
$\partial_\chi\phi$ goes to zero for large $z$, while the components
of the connection $A$ can be chosen to be bounded. 

Let us now investigate the consequence of these two observations.
{}First, one can show that the inverse Nahm
transform yields a traceless connection and a traceless Higgs field
(this is not obvious from their definition). Indeed, if the trace
part of the curvature were nonzero, it would satisfy the Laplace
equation (as a consequence of the Bogomolny equation) and grow at infinity,
in contradiction with the above estimate. Hence the curvature
is traceless and $\Tr\ A$ is a flat connection on $\RR^2\times\SS^1$.
{}Furthermore, $\Tr\ \phi$ must be constant by virtue of the
Bogomolny equation. Now, since
the monopole spectral curve tells us that $\det V(z,2\pi)=1$
everywhere, this means that $\Tr\ \phi=0$ and $\Tr\ A$ has zero
monodromy. Since $\Tr\ A$ is also flat, it must be gauge-equivalent
to zero.

Second, the fact that $\Tr\ V(z,2\pi)$ grows as~(\ref{trgrowth}) at infinity
implies that for large $z$ the eigenvalues of the
Higgs field are 
$$
\pm \frac{k}{2\pi}\log |z|+v+o(1).
$$ 
This proves that the Higgs field has the asymptotics~(\ref{asympt}).
To show that the gauge field has the correct asymptotics, it suffices
to prove that the components of the connection orthogonal to $\phi$
go to zero for large $|z|$, i.e. that the $SU(2)$ monopole
approaches at infinity a $U(1)$ monopole embedded in $SU(2)$.
The argument for this is exactly the same as for the
monopole on $\RR^3$~\cite{JT} (it uses only the fact that at large
distances $||\phi||$ is bounded from below by a strictly positive constant).
In fact,~\cite{JT} proves that the ``nonabelian'' components
of the curvature decay exponentially fast.
{}From the physical point of view this can be explained as follows.
Since $||\phi||\geq 1$ for large enough
$|z|$, the $SU(2)$ gauge group is broken down to $U(1)$,
the Higgs effect makes all the ``nonabelian'' components of the
gauge field massive, and they decay exponentially.

\section{Closing The Circle}\label{sec:Nahmsq}

In this section we prove that the composition of the direct and inverse Nahm transform
takes a periodic monopole to a gauge-equivalent periodic monopole. Together with
the results of Sections~\ref{sec:Nahm} -- \ref{sec:invspec}, this implies that there is
a one-to-one correspondence between the gauge-equivalence classes of periodic $SU(2)$
monopoles with charge $k$ and gauge-equivalence classes of solutions of $U(k)$
Hitchin equations on a cylinder with the asymptotic behavior as described in 
Section~\ref{sec:asymptotics}.
Our proof is modelled on the argument given by Schenk~\cite{Schenk} for 
instantons on a four-torus. Another proof, similar to that given by Donaldson and 
Kronheimer~\cite{DK} for instantons on $T^4$, is sketched in the Appendix.

The direct Nahm transform is defined in terms of square-integrable sections
$\psi_1(x,s),\ldots, \psi_k(x,s)$ of $E\ot S$ which form an orthonormal basis of $\Ker\ \Dd$.
Here $S$ is the spin bundle on $X$, and $\Dd$ is twisted by $s\in\CC$. The sections 
$\psi_1(x,s),\ldots, \psi_k(x,s)$ span a fiber of $\hE$ at a point $s$, so by
combining them
into a matrix $\Psi=\left(\psi_1, \psi_2, \ldots, \psi_k\right)$ we obtain a section
$\Psi(x,s)$ of $\hpi^*(\hE)\ot \pi^*(E\ot S)$. Here $\pi:X\times \hX\ra X$ and
$\hpi:X\times\hX\ra\hX$ are the natural projections. By definition, $\Psi$ satisfies
\begin{equation}
\Dd\Psi=0.
\end{equation}
Since $S$ is trivial and two-dimensional, we can view $\Psi$ as a pair of bundle morphisms 
$(\Psi_1,\Psi_2)$ from $\hpi^*(\hE)$ to $\pi^*(E)$.
(We remind that we have Hermitean
inner products on $E$ and $\hE$ and thus can identify $E$ and $\hE$ with their duals.)

In terms of $\Psi$ the expression for $(\Ah,\phih)$ reads
\begin{equation}
\frac{\partial}{\partial s}-i\Ah_s=\int_X d^3x\ \Tr_{\rm spin}
\Psid(x,s)\frac{\partial}{\partial s}\Psi(x,s),\ \phih=\int_X d^3x\ \Tr_{\rm spin}\Psid(x,s) z
\Psi(x,s).
\end{equation}
We denote by $\frac{\partial}{\partial s}\Psi$ the composition of $\Psi$ and 
$\frac{\partial}{\partial s}$, while the derivative of $\Psi$ with respect to
$s$ will be denoted by $\left[\frac{\partial}{\partial s}, \Psi\right]$.

In order to perform the inverse Nahm transform, we have to find a pair of sections
$\hpsi_1(s,x),\hpsi_2(s,x)$ of $\hpi^*(\hE)\ot \CC^2$ which span $\Ker\ \hD^\dag_{z,\chi}$ for
all $(z,\chi)\in X$.
In other words, if we combine them into a $k\times 2\times 2$ matrix $\Psih=
(\hpsi_1,\hpsi_2)$, $\Psih$ must satisfy
\begin{equation}\label{mario}
\Dhd\Psih=0,
\end{equation}
and, with proper normalization,
\begin{equation}\label{propernorm}
\int_\hX d^2s\ \Tr_{\rm spin}\Psih(s,x)\Psih^\dag(s,x)=1_E.
\end{equation}
Here $1_E$ is the identity endomorphism $E\to E$.

Given $\Psih$, the inverse Nahm transform $(\check{\Ah},\check{\phih})$ is given by
\begin{equation}
d_{\check{\Ah}}=\int_{\Xh} d^2s\ \Psihd(s,x) d_x \Psih(s,x),\ \ \ 
\check{\phih}(s)=\int_{\Xh} d^2s\ \Psihd(s,x)\, r\, \Psih(s,x).
\end{equation}

The difficulty in establishing the equivalence of $(A,\phi)$ and $(\check{\Ah},\check{\phih})$ 
lies in finding
$\Psih(s,x)$ in terms of $\Psi(x,s)$. In the case of the Nahm transform on a four-torus this 
was accomplished by Schenk~\cite{Schenk},
whose results we adapt to the case at hand. Let $\Psih^T$ denote $\Psih$ with the spinor indices 
transposed. We claim that
\begin{equation}\label{llosa}
\Psih^T(s,x)=2\sqrt{2\pi}\int_X d^3y\ \Psid(y,s) (\Dd D)^{-1}(y,x;s) e^{-i\chi_x t},
\end{equation}
where $t={\rm Im} s$, as usual. In what follows it will be convenient to regard $x,y\in X$
as continuous labels and think of $\Psi$ as an object with one continuous and three
discrete labels. Integration over $x$ is then regarded as a summation over a continuous
label and is not shown explicitly. The dependence on $s$ will not be shown explicitly
either. In this shortened notation Eq.~(\ref{llosa})
takes the form
$$
\Psih^T=2\sqrt{2\pi}\ \Psid (\Dd D)^{-1} e^{-i\chi t}.
$$

{}The first thing to check is whether $\Psih$ is a well-defined section of $\hE$, that is,
whether $\Psih(s+i,x)=\Psih(s,x)$. 
Since the twisted derivative along $\chi$  is given by
$\left(\partial_{\chi}-i A_{\chi}+i t\right)$, we see that $\Psi(x,s+i)$ is 
related to $e^{-i\chi} \Psi(x,s)$ by a $U(k)$ gauge transformation. By making an 
$s$-dependent change of
basis in $\Ker\ D^\dag$, we can always ensure that $\Psi(x,s+i)=e^{-i\chi}\Psi(x,s)$.
Furthermore, we have
$$\G(x,y;s+i)=e^{-i\chi_x}\G(x,y; s) e^{i\chi_y}.$$
It follows that $\Psih(s+i,x)=\Psih(s,x)$.

In terms of $\sigma_{\pm}=\sigma_1\pm i\sigma_2$ and $p_{\pm}=(1\pm \sigma_3)/2$
the  untwisted operators $\Dh$ and $\Dhd$ take the following form
$$
\Dh=
\begin{pmatrix}
\partial_{\Ah},&\bar{\partial}_{\Ah},&\phih,&\phihd
\end{pmatrix}
\begin{pmatrix} 
\sigma_+ \\ 
\sigma_- \\ 
-p_+\\ 
-p_-
\end{pmatrix}
$$
and
\begin{equation}
\Dhd=(-1)\begin{pmatrix}\partial_{\Ah},&\bar{\partial}_{\Ah},&\phih,&\phihd\end{pmatrix}\begin{pmatrix}
\sigma_+\\ 
\sigma_-\\
p_-\\
p_+
\end{pmatrix}.
\end{equation}
The statement that $\Psih(s,x)$ given by (\ref{llosa}) satisfies (\ref{mario}) is equivalent to 
the following identity
\begin{equation}\label{identdirac}
\left(\Psid\left(\left[\frac{\partial}{\partial s},\Psi\right],
\left[\frac{\partial}{\partial \bar{s}},\Psi\right],z\Psi,\bz\Psi\right) 
\Psid (\Dd D)^{-1}-
\Psid (\Dd D)^{-1} (0,0,z,\bz) \right) 
\begin{pmatrix}
\sigma_+^T\\ 
\sigma_-^T\\
p_-^T\\
p_+^T
\end{pmatrix}=0.
\end{equation}
We remind that $z$ stands for an operator of multiplication by $x_1+ix_2$, or
equivalently for an integral operator with a kernel $(x_1+ix_2) \delta(x,y)$.

Making use of the identities
\begin{equation}
\Psi\Psid=1- D (\Dd D)^{-1} \Dd,
\end{equation}
\begin{eqnarray}\nn
\left[\frac{\partial}{\partial s}, (\Dd D)^{-1}\right]=\G (-p_- D-\Dd p_+) \G,\\
\label{identities} \left[\frac{\partial}{\partial \bar{s}}, (\Dd D)^{-1}\right]=\G (-p_+ D-\Dd p_-) \G, \\
\left[z,\G\right]=\G (-\sigma_+ D + \Dd \sigma_+) \G,\nn \\
\left[\bz, \G\right]=\G (-\sigma_- D + \Dd \sigma_-) \G,\nn 
\end{eqnarray}
and $\sigma_{\pm}^T=\sigma_{\mp}$ and $p_{\pm}^T=p_{\pm}$, one can see that 
Eq.~(\ref{identdirac}) is a consequence of  
a matrix identity
\begin{equation}\nn
\left((p_-,p_+,\sigma_+,\sigma_-) D+( p_+,p_-,-\sigma_+,-\sigma_-) \Dd
+\Dd (p_+,p_-,-\sigma_+,-\sigma_-) \right) 
\begin{pmatrix}
\sigma_-\\
\sigma_+\\
p_-\\ 
p_+\\
\end{pmatrix}=0,
\end{equation}
which can be readily verified.
Thus we conclude that $\Psih$ defined by Eq.~(\ref{llosa}) indeed solves $\Dhd\Psih=0$.

Next we verify Eq.~(\ref{propernorm}). With the help of Eqs.~(\ref{identities})
we find
\begin{eqnarray}
&&\Tr_{\rm spin}\G\left(1-D\G\Dd\right)\G= \\
&&=-\frac{1}{8}\,\Tr_{\rm spin}\left(4\partial_{\bar{s}}\partial_s\G-\left[\bar{z}, 
\left[z,\G\right]\right]\right).\nn
\end{eqnarray}
Combining this identity with the the formula for $\Psih$, we obtain
\begin{eqnarray}\label{norm}
&&\int_{\Xh} d^2s\ \Tr_{\rm spin} \Psihd(x_1,s)\Psih(x_2,s)=\nn \\ 
&&=-\pi\ \Tr_{\rm spin}\int_{\Xh} d^2s\ e^{i t(\chi_1-\chi_2)}
\left(4\partial_{\bar{s}}\partial_s-(\bar{z_1}-\bar{z_2})(z_1-z_2)\right)\G(x_1,x_2;s).
\end{eqnarray}

Integrating by parts and considering the limit of $x_2$ approaching $x_1$, we
see that in this limit the integral is dominated by the region of large $r={\rm Re}\ s$. 
Thus to estimate the integral it is sufficient to consider the large $r$ limit,
where $\G$ reduces to the Green's function of the operator $-\nabla^2+r^2$. 
We conclude that in the limit $x_1\ra x_2$ the right-hand side of Eq.~(\ref{norm}) reduces to
$$
-2\pi 1_E\lim_{x_1\ra x_2}\int_{-\infty}^{+\infty} dr \left(-|x_1-x_2|^2\right) 
\frac{e^{-|r| |x_1-x_2|}}{4\pi |x_1-x_2|}.
$$
Performing the integral over $r$, we get~(\ref{propernorm}).

Now, having constructed $\Psih$, one can find the result of the inverse Nahm transform. 
Let us start with  a
few useful identities valid for any smooth section $\Xi(s)$ of $\hE$ which decays rapidly
as $|s|\rightarrow\infty$ together with all its derivatives (i.e. belongs to the Schwarz space):
\begin{eqnarray}
\int_{\Xh} d^2s\ \Psi(x,s)\ \partial_{\Ah+\ha} \Xi(s)&=&\int_{\Xh} d^2s\ D \G p_- \Psi\  \Xi(s),\nn \\
\int_{\Xh} d^2s\ \Psi(x,s)\ \bar{\partial}_{\Ah+\ha} \Xi(s)&=&\int_{\Xh} d^2s\ D \G p_+ \Psi\  \Xi(s),\nn \\
\int_{\Xh} d^2s \left(\Psi(x,s) \phih-z\Psi(x,s)\right) \Xi(s)&=&\int_{\Xh} d^2s\ D \G \sigma_+ \Psi\  \Xi(s),\nn \\
\int_{\Xh} d^2s \left(\Psi(x,s) \phihd-\bz\Psi(x,s)\right) \Xi(s)&=&\int_{\Xh} d^2s\ D\G \sigma_- \Psi\  \Xi(s).\nn
\end{eqnarray}
Now substitute into these formulas $\Xi(s)=e^{it\chi'}\Psih^T(s,x')$, set $x'=x$,
multiply them from the right by
$\sigma_{-}, \sigma_+, p_-,$ and $p_{+}$, respectively, and sum them up.
The left-hand side of the resulting identity will be proportional to $\left(\Dd\Psih\right)^T$ 
and therefore will vanish. Thus we get
\begin{equation}
\int_{\Xh} d^2s\ D \G \left(p_-,p_+,\sigma_+,\sigma_-\right)\Psi 
e^{i\chi t} \Psih^T\begin{pmatrix} \sigma_-\\ \sigma_+\\ p_-\\ p_+\end{pmatrix}=0.
\end{equation}
Substituting $2\sqrt{2\pi}e^{i\chi t}\G\Psi =\left(\Psih^T\right)^{\dagger}$ and using an identity 
\begin{equation}
(p_-,p_+,\sigma_+,\sigma_-) M \begin{pmatrix}\sigma_-\\ \sigma_+\\ p_-\\ 
p_+\end{pmatrix}=(\sigma_+ +\sigma_-) \Tr_{\rm spin} M
\end{equation}
valid for any $2\times 2$ matrix $M$,
we are left with
\begin{equation}\label{comp}
\int_{\Xh} d^2s\ e^{i\chi t}D e^{-i\chi t} (\sigma_+ +\sigma_-)\Tr_{\rm spin} \left(\Psih^T\right)^{\dagger}\Psih^T=0.
\end{equation}
This operator equation in spin space can be rewritten as four ``scalar'' equations which
express the coincidence of $(A,\phi)$ and $(\check{\hA},\check{\hphi})$. 
For example, the vanishing of the coefficient of $p_+$ in Eq.~(\ref{comp}) implies
\begin{equation}
\int_{\Xh} d^2s \left(\left(\frac{\partial}{\partial z}-
i{A_z}\right)\Psihd_1(s,x)\right)\Psih_1(s,x)+\left(\left(\frac{\partial}{\partial z}-i{A_z}\right)
\Psihd_2(s,x)\right)\Psih_2(s,x)=0,
\end{equation}
where $\Psih_1$ and $\Psih_2$ are the two spinor components of $\Psih$. This equation
simply says that $A_z=\check{\Ah}_z$.
Writing out the coefficient of $\sigma_+$, we obtain
\begin{multline}
\int_{\Xh} d^2s \left(\left(\frac{\partial}{\partial\chi}-i A_\chi-\phi+r\right)\Psihd_1(s,x)\right)
\Psih_1(s,x)\\
+\left(\left(\frac{\partial}{\partial\chi}-i A_\chi-\phi+r\right)\Psihd_2(s,x)\right)
\Psih_2(s,x)=0,
\end{multline}
which implies $A_{\chi}=\check{\Ah}_{\chi}$ and $\phi=\check{\phih}$.
This completes the proof.

\section{Remarks On The Existence Of Periodic Monopoles}\label{sec:existence}

It is intuitively plausible that periodic monopoles exist for all $k>0$.\footnote{From the string
theory point of view, periodic $SU(2)$ monopoles can be identified with D4 branes
suspended between two parallel NS5 branes, with one direction common to D4-branes and
NS5-branes compactified on a circle (see section~\ref{sec:gaugetheory} for details).
This brane configuration surely exists, so one is tempted to dismiss the question
of the existence of periodic monopoles as trivial. But it is far from obvious that
suspended D4-branes are represented by nonsingular field configurations on the NS5-branes
after a T-duality along the compact direction.}
In fact, when the parameter
$v$ in~(\ref{phiD}) is large, one can propose a simple way of constructing approximate solutions
of Bogomolny equations on $\RR^2\times S^1$. One considers a charge $k$ $SU(2)$ monopole on 
$\RR^3$ located
near the origin of $\RR^3$. It approaches a charge $k$ Dirac monopole 
exponentially fast at distances larger than $1/v$~\cite{JT}. Then one can patch it with a periodic
Dirac monopole solution of Section~\ref{sec:intro} at distances larger than $1/v$ but smaller than $1$,
and obtain an accurate approximation to a nonabelian periodic monopole. 

To prove the existence of periodic monopoles for all $v$ and $k$ it is easier to use the
correspondence between periodic monopoles and solutions of Hitchin equations on a cylinder
established above.  For $k=1$ one can write down explicitly a family of solutions of Hitchin
equations with required asymptotics:  
\begin{equation}\label{solutionkone} 
\hphi(s)= e^{2\pi (\ups+s)}+ e^{2\pi (\ups-s)}+c,\quad c\in\CC, 
\qquad A=\beta dt,\quad \beta\in \RR/(2\pi\ZZ).
\end{equation} 
This proves that a periodic monopole of charge $1$ exists.  Moreover, it is
easy to see that any solution of $U(1)$ Hitchin equations with the boundary conditions described
in section~\ref{sec:asymptotics} is gauge-equivalent to~(\ref{solutionkone}).  Thus a
periodic monopole with $k=1$ has three real moduli (${\rm Re}\ c, {\rm Im}\ c, \beta$). 
(Caution: we do not claim that there is a natural metric on this moduli space, and in fact
we will see below that this is not true.)  They
arise from the translational invariance of the Bogomolny equations and parameterize
$\RR^2\times\SS^1=X$.  We may regard them as describing the location of the monopole on $X$.

{}For $k>1$ finding solutions of Hitchin equations is harder, and we do not have a satisfactory
proof of their existence. Below we merely sketch a possible approach to the proof based on
the holomorphic description of solutions of Hitchin equations.
The idea of the holomorphic approach is familiar
to physicists in the guise of the following principle: the space of solutions of
D and F-flatness conditions modulo a compact gauge group is the same as the space of 
solutions 
of the F-flatness conditions modulo the complexified gauge group (this principle is often 
referred to as the Luty-Taylor theorem~\cite{LT}). 

Let us apply this principle to our problem.
The ``complex'' Hitchin equation is invariant with respect to
the complexified gauge transformations, i.e. gauge transformations which are 
$GL(k,\CC)$-valued.
The ``real'' Hitchin equation is invariant only with respect to $U(k)$ gauge 
transformations.
Thus from the physical point of view, the ``real' 
and ``complex'' Hitchin equations play the role of the D-flatness and F-flatness 
conditions, 
respectively, and it is natural to consider the space of solutions of the ``complex'' 
Hitchin equation
modulo the complexified gauge group. The Hermitian inner product on the bundle $\hE$ then 
plays no role, and all we have is a holomorphic bundle over $\hX\cong\CC^*$. 
The ``complex'' Hitchin equation
says that $\hphi$ is a holomorphic section of $\End(\hE)$. Such a pair, a holomorphic bundle 
$\hE$ on $\hX$
and a holomorphic section of $\End(\hE)\ot \Omega_\hX$, is called a Higgs bundle. 
Obviously, we have a forgetful map from the moduli space of solutions of the full 
Hitchin equations 
to the moduli space of Higgs bundles. 

It is very easy to construct Higgs bundles on $\hX$, and in the next section we will give
a rather explicit description of their moduli space. Thus if the above-mentioned map is surjective,
the existence of solutions of Hitchin equations will be established.

Let us explain why it is plausible that every suitable Higgs bundle comes from a solution of
Hitchin equations. In the case of Hitchin equations on a compact Riemann surface, 
one can prove that any {\it stable} Higgs bundle is related by a $GL(k,\CC)$ gauge transformation
to a solution of Hitchin equations. The role of the stability condition is to ensure that the 
complexified gauge group acts freely on the Higgs bundles. One may also consider solutions of
Hitchin equations on a punctured Riemann surface with ``tame'' singularities at the 
punctures~\cite{Simpson,Konno}.
(``Tame'' means that the eigenvalues of the Higgs field grow at most as $1/r$ as one approaches
the puncture.) The corresponding holomorphic object is a Higgs bundle on the punctured Riemann
surface whose Higgs field has simple poles at the punctures. Again there is a stability condition
on the Higgs bundle which ensures that the complexified gauge group acts freely on its orbit,
and any stable Higgs bundle on a punctured Riemann surface comes from a solution of Hitchin equations
with tame singularities~\cite{Simpson,Konno}. 

In our case the Higgs field is not ``tame'' at infinity. To see this, let us make a conformal 
transformation $w=\exp(2\pi s)$ which maps the cylinder to $\CC^*$. Keeping in mind that the 
Higgs field is a section of $\End(\hE)\ot\Omega^{1,0}$, we see that its eigenvalues near $w=0$ 
behave as $|w|^{-1-1/k}$, i.e. the singularity is not ``tame.'' A similar problem occurs at 
$w=\infty$. This means that we cannot use the results of~\cite{Simpson, Konno}. Still, the
above discussion suggests that the important thing is for complexified gauge transformations
to act freely on the Higgs bundles. It appears that this condition is always satisfied if the spectral
curve of the Higgs bundle is given by
\begin{equation}\label{Higgscurve}
z^k+a_1 z^{k-1}+\ldots+a_k- 2 e^{-2\pi\ups}\cosh(2\pi\,s)=0.
\end{equation}
Indeed, if there were a $GL(k,\CC)$ transformation
which would leave our Higgs bundle $(\hE,\hphi)$ invariant, this would mean that $\hE$ has a 
rank one holomorphic subbundle invariant with respect to $\hphi$.
However, this is clearly not true for large $|{\rm Re} s|$ because of the 
asymptotic behavior of the eigenvalues of $\hphi$: they are all distinct and cyclically permuted
as one goes around the circumference of the cylinder. Thus it seems plausible that any Higgs
bundle whose spectral curve has the form~(\ref{Higgscurve}) is related by a $GL(k,\CC)$ 
transformation to a solution of Hitchin equations with required asymptotics.

One could ask if there could be a one-to-one correspondence between solutions of Hitchin
equations on a cylinder modulo $U(k)$ gauge transformations and holomorphic Higgs bundles 
with the spectral curve~(\ref{Higgscurve}) modulo $GL(k,\CC)$ gauge transformations.
This would be the analogue of the Luty-Taylor theorem for Hitchin equations on a cylinder. 
If the question is posed this way,
the answer is negative. Indeed, we already saw that rank one solutions of Hitchin equations 
are
parameterized by a complex number $c$ which describes the Higgs field, and a real number 
$\beta$ which parameterizes the monodromy of $\hA$ around the circumference of the cylinder.
On the other hand, the bundle $\hE$ is holomorphically trivial, so all the information about
the holomorphic Higgs bundle is described by $c$. In other words, the information about the
monodromy of $\hA$ is lost in the holomorphic picture. 

In the case of Higgs bundles with ``tame'' singularities, the situation is similar: the 
information
about the monodromy of $\hA$ around the punctures is lost upon passing to a Higgs bundle. But
there is a way to fix this: one has to consider Higgs bundles with ``parabolic structure''
at the punctures~\cite{Simpson}. Parabolic structure essentially encodes the conjugacy class
of the monodromy. In our case it is reasonable to conjecture that what is missing in the
naive holomorphic description is precisely the information about the monodromy of $\hA$ 
at infinity.
In fact, we have an a priori knowledge (see Section~\ref{sec:asymptotics})
that the monodromy is given by $\exp(i\beta) V^{\pm 1}$, where $V$ is the ``shift''
matrix~(\ref{shift}), and $\beta\in \RR/(2\pi\ZZ)$. Thus the holomorphic 
description misses one real parameter $\beta\in \RR/(2\pi\ZZ)$. 

We are led to the following conjecture. Let ${\cal M}_{Hi,k}$ be the moduli space of
solutions of $U(k)$ Hitchin equations on a cylinder with asymptotics~(\ref{asympttwo}). Let 
${\cal M}_{HB,k}$ be the moduli space of holomorphic Higgs bundles on $\CC^*$
whose spectral curve has the form~(\ref{Higgscurve}). The forgetful map from ${\cal M}_{Hi,k}$
to ${\cal M}_{HB,k}$ is surjective, and moreover is a fiber bundle with fiber $\SS^1$.

In the next section we will test this conjecture by computing the dimension of ${\cal M}_{HB,k}$
and comparing with expectations from ${\cal N}=2$ super-Yang-Mills.

\section{The Moduli Space Of Periodic Monopoles}\label{sec:modulispace}

In this section we describe the moduli space ${\cal M}_{HB,k}$ of solutions of the
complex Hitchin equation on $\RR\times\SS^1$ with the spectral curve (\ref{Higgscurve}). 
Assuming that the analogue of the Luty-Taylor theorem formulated
in the previous section is true, the moduli space of charge $k$ periodic monopoles is fibered over
${\cal M}_{HB,k}$ with fiber $\SS^1$. As explained below, there is an alternative
way to view the relation between ${\cal M}_{HB,k}$ and periodic monopoles: a certain
submanifold in ${\cal M}_{HB,k}$ of complex codimension one coincides with the {\it centered} moduli
space of charge $k$ periodic monopoles. We compare our results with the expectations from 
string theory and discuss the existence of a hyperk\"ahler metric on the centered moduli space of 
periodic monopoles.

We already know how to associate a spectral curve $\C\in \CC^*\times \CC^*$ and a coherent sheaf 
$N$ on it to every solution of the complex Hitchin equation. For a generic
solution, the curve $\C$ is nonsingular, and $N$ is a line bundle 
(see Section~(\ref{sec:Nahm})). The curve
has the form
$$
w^2-w f(z)+1=0,
$$ 
where $f(z)$ is a polynomial of degree $k$ whose leading coefficient is a known constant. Thus to
specify the polynomial $f(z)$ we need to specify its $k$ coefficients $a_1,\ldots,a_k$.

{}For $k=1$ the spectral curve is rational, and its compactification is
a $\PP^1$. To understand what happens for $k>1$, it is convenient to rewrite the
equation of the curve $\C$ in the form
\begin{equation}\label{newform}
\tw^2=\frac{1}{4} f(z)^2-1,
\end{equation}
where $\tw=w-f(z)/2$. This equation implies that the compactification of $\C$ is a hyperelliptic
curve in $\PP^2$. It is well known that a hyperelliptic curve in $\PP^2$
has singularities, in this case over the point $z=\infty$. Its desingularization has genus $k-1$ and
will be denoted $\tC$. 
The pull-back of the line bundle $N$ to $\tC$ will be denoted by the same letter $N$.
Since $\hE$ is a trivial bundle, $N$ has zero degree.
The moduli space of line bundles over $\tC$ with fixed degree is simply the Jacobian of 
$\tC$, which is an Abelian variety of dimension $k-1$. 

Conversely, starting from a hyperelliptic curve $\tC$ and a line bundle
over it, we can reconstruct the solution of the complex Hitchin equation. The bundle $\hE$
over $\CC^*$ is obtained by pushing forward $N$ with respect to the 
projection $(w,z)\mapsto w$. The Higgs field $\hphi\in \Gamma(\End(\hE))$ is defined
as follows:
$$
\hphi: v_{w,z}\mapsto z\ v_{w,z}.
$$

Thus we obtain the following description of the moduli space of ${\cal M}_{HB,k}$ valid in an 
open set: it is the space of pairs $(\tC,N)$, where $\tC$ is a hyperelliptic curve of genus $k-1$
and $N$ is a degree $0$ line bundle on it. Hence ${\cal M}_{HB,k}$
is a complex manifold of dimension $2k-1$ fibered over $\CC^k$ by Abelian varieties of 
dimension $k-1$.

It follows that the moduli space ${\cal M}_{Hi,k}$, and therefore
the moduli space of periodic monopoles of charge $k$,
has real dimension $4k-1$. This coincides with the dimension of the moduli space of
$SU(2)$ monopoles of charge $k$ on $\RR^3$. 
{}Furthermore, string theory predicts that the {\it centered} moduli space
of periodic monopoles has dimension $4k-4$ (see section~\ref{sec:gaugetheory}). 
Centering the monopole amounts to
setting $\beta=0$, and $a_1=0$, where $a_1$ is the coefficient of $z^{k-1}$ in $f(z)$.
Indeed, we already explained in Section~\ref{sec:specS} that the positions of the 
constituent charge $1$ monopoles on $\RR^2$
are the roots of the equation $f(z)=2$, so setting $a_1=0$ has the effect of 
making the center-of-mass of the monopole located at $z=0$. It is also easy to check
that a translation along $\SS^1$ has the effect of shifting $\beta$. Thus the centered
moduli space of periodic monopoles of charge $k$ is a hypersurface in ${\cal M}_{HB,k}$ given
by the equation $a_1=0$. It has complex dimension $2k-2$, in agreement with
string theory predictions. This lends support to the conjectured correspondence
between solutions of Hitchin equations on a cylinder and a special class of Higgs bundles.

Moreover, one expects on physical grounds that the moduli space of the ${\cal N}=2$ 
super Yang-Mills compactified on a circle has a distinguished complex structure in which it is a complex manifold
fibered over $\CC^{k-1}$ by Abelian varieties of dimension $k-1$~\cite{SW,DonagiWitten}.
We saw above that this is indeed true. 

Let us now turn to the issue of the hyperk\"ahler metric on the moduli space of periodic
monopoles. Supersymmetry implies that the Coulomb branch of the ${\cal N}=2$ $SU(k)$
super-Yang-Mills theory compactified on a circle must be a complete hyperk\"ahler 
manifold~\cite{SWthree}, so
we expect that the hyperk\"ahler metric exists for the {\it centered} moduli space.
In contrast to monopoles on $\RR^3$, we do not expect to have a well-defined metric
on the uncentered moduli space. The reason for this is that the uncentered monopoles
would correspond to a $U(k)$ gauge theory in $d=4$, but the latter does not make sense as
a quantum theory because it is not asymptotically free.

We can also explain this in a purely classical way, which does not involve quantum
${\cal N}=2$ super-Yang-Mills theory. The difference between the centered and the uncentered
moduli spaces is that in the former case we mod out by translations of $\RR^2\times\SS^1$,
while in the latter case we don't. The reason why one needs to divide by the translations
group to get a well-defined metric is that the tangent vectors to the moduli space corresponding
to the translations on $\RR^2$ are not normalizable, i.e. their $L^2$ norm diverges.
This tangent vector is given by $(\delta A,\delta \phi)=(\partial_z A,\partial_z\phi).$ 
According to~(\ref{phiD}), $\partial_z\phi$ decays only as $1/z$, therefore the 
$L^2$ norm of this tangent vector is logarithmically divergent.

The above arguments demonstrate that there is no well-defined metric on the uncentered moduli space,
but they do not prove that there is one on the centered moduli space. This can be argued as follows. 
As explained in Section~\ref{sec:asymptotics}, for large $|z|$ a nonabelian  periodic
monopole is exponentially close to a periodic Dirac monopole embedded in $SU(2)$. Thus to count 
$L^2$ deformations it is
sufficient to use the abelian asymptotics~(\ref{phiD},\ref{AD}). Then it is easy to see that changing
the locations of monopoles while keeping their center-of-mass fixed changes the Higgs field
only by terms which decay as $1/|z|^2$ (this is essentially multipole expansion).
Thus all such deformations have finite $L^2$ norm. There are $3k-3$ such tangent vectors.
Using the quaternionic structure of the tangent space (see below), one can show that the 
remaining $k-1$ tangent vectors are also normalizable.

{}From the mathematical point of view, it may be easier to count $L^2$ deformations
in the Nahm-transformed picture. 
Note that setting $\beta=0$, $a_1=0$ amounts to setting
$\Tr\ \hphi=0$ and passing from the $U(k)$ to $SU(k)$ Hitchin equations.
$SU(k)$ Hitchin equations may be regarded as hyperk\"ahler moment map equations for the 
action of the $SU(k)$ gauge group on the cotangent bundle of the space of $SU(k)$ connections
on $\RR\times\SS^1$~\cite{HitchinSpec}. Formally, the hyperk\"ahler quotient 
construction~\cite{HKLR} implies that the moduli
space of $SU(k)$ Hitchin equations has a hyperk\"ahler metric. In order to
prove the existence of a hyperk\"ahler metric on the centered moduli space,
it is sufficient to show that the space of $L^2$ deformations of $SU(k)$ Hitchin
equations on a cylinder has the expected dimension $4k-4$.

The properties of the hyperk\"ahler metric on the centered moduli
space of periodic monopoles will be discussed elsewhere~\cite{usthree}. One thing is
clear though: in the limit when the circumference of $\SS^1$ goes to infinity, the centered
moduli space of periodic monopoles smoothly goes over to the centered moduli space of 
monopoles on $\RR^3$. In particular, the metric on the centered moduli space 
of a charge $2$ periodic monopole is a deformation of the Atiyah-Hitchin metric. 
It would be very interesting to find the explicit form of this metric. On physical
grounds, we expect that it is hyperk\"ahler and asymptotically locally flat.
But unlike the Atiyah-Hitchin metric, which has an $SU(2)$ isometry, the new metric seems to
have no continuous isometries.

\section*{Acknowledgments}
We are grateful to Nigel Hitchin for a very helpful conversation
concerning the definition of the monopole spectral data, and
to Dmitri Orlov and  Marcos Jardim for discussions. We also wish to thank the 
organizers of the workshop ``The Geometry and Physics of Monopoles,'' Edinburgh,
August-September 1999, for creating a very stimulating atmosphere during the meeting
and for providing us with an opportunity to present a preliminary version of this work.
The work of S.Ch. was supported in part by NSF grant PHY9819686. The work of A.K. was
supported in part by a DOE grant DE-FG02-90ER4054442.

\section*{Appendix}

In section~\ref{sec:Nahmsq} we proved that the composition of the Nahm transform 
of Section~\ref{sec:Nahm} and the inverse Nahm transform of Section~\ref{sec:Inverse} is 
the identity map on the gauge-equivalence classes of periodic monopole configurations. 
For the sake of completeness, we present 
here an outline of a cohomological proof of this fact in the spirit of reference~\cite{DK}.
Both the direct and inverse Nahm transforms,
as well as a map identifying the result of the composition of the two transforms
with the initial configuration, will emerge from the spectral sequence of a double complex.

Nahm transform, as described in subsection~\ref{subsec:Coh}, is given in terms of the cohomology of the 
following complex:
\begin{equation}\label{straight}
0\rightarrow \Lambda^{0,0}(X,E)\xrightarrow{\bD_0}\Lambda^{0,1}(X,E)
\xrightarrow{\bD_1}\Lambda^{0,2}(X,E)\rightarrow 0.
\end{equation}
The differentials $\bD_p,\ p=0,1,$ here are twisted by $\hx\in\hX$.
In the trivialization of the Poincar\/{e} bundle defined in Section~\ref{sec:invspec} 
the operators $\bD_0$ and $\bD_1$  are given by:
\begin{equation}
\bD_p=\dbz\wedge 2 (\frac{\partial}{\partial{\bar{z}}}-iA_{\bz})+d\chi\wedge
(\frac{\partial}{\partial{\chi}}-iA_{\chi} -\phi+r).
\end{equation}

Consider a trivial bundle 
$\Ebl\rightarrow X\times\hX$, such that its restriction to $X\times\hx$
is $E$ and restriction to $x\times\hX$ is a trivial bundle with 
fiber $E\vert_{x}$. For each $\hx\in\hX$ it has an action 
of the operator $\bD$ twisted by $\hx$. 
Let $\Omega^p$ denote the sheaf of $\Ebl$-valued rapidly decaying $p$-forms spanned by the
differentials $d\chi$ and 
$d\bar{z}$ with coefficients depending on $x$ and $\hx$. Here by ``rapidly decaying'' we
mean rapidly decaying both for large $|z|$ and large ${\rm Re}\ s$. 
Consider a double complex
\begin{equation}\label{co}
\xymatrix{
&&\Omega^{2}\ar[r]^-{\delta_0} &\Omega^{2}\oplus\Omega^2\ar[r]^-{\delta_1} &\Omega^{2}\\
\CO^{p,q}: &&\Omega^{1}\ar[r]^-{\delta_0}\ar[u]^{\bD_1} &\Omega^{1}\oplus\Omega^1\ar[r]^-{\delta_1}\ar[u]^{\bD_1} &\Omega^{1}\ar[u]^{\bD_1}\\
&&\Omega^{0}\ar[r]^-{\delta_0}\ar[u]^{\bD_0} &\Omega^{0}\oplus\Omega^0\ar[r]^-{\delta_1}\ar[u]^{\bD_0} &\Omega^{0}\ar[u]^{\bD_0},
}
\end{equation}
where $\delta_0$ and $\delta_1$ act as follows:
\begin{equation}
\delta_0: f\rightarrow\left(
        \begin{array}{c} -z f\\ 2\left(\frac{\partial}{\partial\bar{s}}+\frac{1}{2}\chi\right) f
        \end{array}   \right),\ \ 
\delta_1:\left(
        \begin{array}{c} g_0\\g_1 \end{array}
          \right)\rightarrow \left(-2\left(\frac{\partial}{\partial\bar{s}}+\frac{1}{2}\chi\right) g_0-z g_1\right). 
\end{equation}
It is easy to check that $\bD$ and $\delta$ commute. The zeroth and second cohomology 
of $\bD$ vanish, and the first cohomology yields $\hE$. 
Thus the cohomology of columns with respect
to $\bD$ is:
\begin{equation}
\xymatrix{
&&0\ar[r]&0\ar[r]&0\\
\tE_1^{p,q}:&&\hLambda^0\ar[r]^-{\hdelta_0}&
\hLambda^0\oplus\hLambda^1\ar[r]^-{\hdelta_1}&\hLambda^1\\
&&0\ar[r]&0\ar[r]&0,
}
\end{equation}
which contains exactly the sequence~(\ref{thehat}) defining the inverse Nahm transform. Thus
on the second level the sequence degenerates to
\begin{equation}\label{hatE}
\xymatrix{
&&0&0&0\\
\tE_2^{p,q}:&&H^0_{\hdelta}(\hX,\hE)&
\check{\hE}=H^1_{\hdelta}(\hX,\hE)& H^2_{\hdelta}(\hX,\hE)\\
&&0&0&0.
}
\end{equation}

Computation of the other spectral sequence $E^{p,q}$ is exactly analogous to that 
of~\cite{DK}, the result being
\begin{equation}
\xymatrix{
&&0&0&0\\
E_2^{p,q}:&&0&0&0\\
&&0&0&E\vert_{x=0}.
}
\end{equation}
Comparing the total cohomologies of the double complex~(\ref{co})
\begin{equation}
\oplus_{p+q=n} E^{p,q}_{\infty}=\oplus_{p+q=n} \tE^{p,q}_{\infty},
\end{equation}
we conclude that $E|_{x=0}=\check{\hE}|_{x=0}$.

Chasing the spectral sequence we can obtain the isomorphism 
$\omega: \check{\hE}|_{x=0}\rightarrow E|_{x=0}$
explicitly. Namely, an element of $\check{\hE}|_{x=0}$ can be represented by 
$\alpha\in\Omega^{1}\oplus\Omega^1$ harmonic
with respect to $\bD$ and such that $\delta_1\alpha=\bD\beta$ for some $\beta\in\Omega^{0}$.
The isomorphism $\omega$ takes $\alpha$ to the value of $\beta$ at $\hx=0$ integrated over 
$X\times \{0\}$.
The equation for $\beta$ is solved by 
$\beta=\left(\bD^*\bD\right)^{-1}\bD^*\delta_1\alpha,$ thus
\begin{equation}\label{omega}
\omega:\alpha\mapsto\int_{X\times\{0\}}dx \left(\bD^*\bD\right)^{-1}\left[\bD^*, \delta_1\right]\alpha.
\end{equation}
Note that $\left(\bD^*\bD\right)^{-1}$ is proportional to the Green's function $\G$, while 
the commutator $\left[\bD^*, \delta_1\right]$ has a simple form: it maps
$(g_0, g_1)\in\CO^{1,1}=\Omega^1\oplus\Omega^1$ to $2 (g_{0,\chi}+g_{1,\bar{z}})$.

The point $x=0$ was not distinguished in any natural way, and twisting the vertical operator 
$\delta$ of the
double complex~(\ref{co}) by $(-x_0)\in X$ and computing the spectral sequence would lead to an isomorphism
$\omega: \check{\hE}|_{x=x_0}\rightarrow E|_{x=x_0}$. Therefore the isomorphism $\omega$ is an isomorphism
of bundles on $X$. Using the above explicit formula for $\omega$, it can be checked
that it commutes with the differential $\bD$. This shows that 
$\check{\hA}_{\bar{z}}=A_z$ and 
$(i\check{\hA}_{\chi}+\check{\hat{\phi}})=(iA_\chi+\phi)$.

In order to conclude that the isomorphism $\omega$ takes the original monopole data 
$(A, \phi)$ to 
$(\check{\hA}, \check{\hat{\phi}})$ we need to say a few words regarding the naturalness of 
the above 
construction. The way to present the direct (as well as inverse) Nahm transform in 
cohomological terms is not unique. For example, we could have replaced $\bD$ with a differential
\begin{equation}
(dx_1-i d\chi)\wedge\left(\frac{\partial}{\partial x_1}-i A_1-i\left(\frac{\partial}{\partial \chi}-i A_{\chi}\right)\right)+
dx_2\left(\frac{\partial}{\partial x_1}-i A_2+(\phi-r)\right),
\end{equation}
and modified the cohomological construction accordingly. This amounts to identifying
$X\cong \RR\times \CC^*$ instead of $X\cong \CC\times \SS^1$. This arbitrariness 
is exactly the same as the arbitrariness in the choice of complex structure in the twistor 
description of monopoles, as well as in the
discussion of Higgs bundles and Nonabelian Cohomology in \cite{Simpson}. As in the case 
of a four-torus \cite{DK}, we could have constructed an isomorphism 
$\eta: \check{\hat{E}}\rightarrow E$ preserving an appropriate differential
for each choice of the identification $X\cong \RR\times \CC^*$.
We would have discovered then that the isomorphism $\eta$ is always given by the formula 
(\ref{omega}) and therefore coincides
with $\omega$. Therefore, $\omega$ maps $(\check{\hA},\check{\hphi})$ to
$(A,\phi)$.

%%%%%%%%%%%%%%%%%%%%%%%%%%%%%%%%%%%%%%%%%%%%%%%%%%%%%%%%%%
 
\end{document}